\documentclass{IEEEtran}

\usepackage{enumitem,kantlipsum}
\usepackage{graphicx}
\usepackage{epstopdf}
\usepackage{amssymb}
\usepackage{amsmath}
\usepackage{subfigure}
\usepackage{epsfig}
\usepackage{color}
\usepackage{enumitem}
\usepackage{float}
\usepackage{soul}
\usepackage{wrapfig}
\usepackage{arydshln}
\usepackage{tablefootnote}

\usepackage{amsthm}

\def\0{{\bf 0}}
\def\1{{\bf 1}}

\def\beq{\begin{equation*}}
    \def\eeq{\end{equation*}}
\def\bql{\begin{equation}}
    \def\eql{\end{equation}}
\def\bqn{\begin{eqnarray*}}
    \def\eqn{\end{eqnarray*}}
\def\bnl{\begin{eqnarray}}
    \def\enl{\end{eqnarray}}
\def\bma{\begin{bmatrix}}
    \def\ema{\end{bmatrix}}
\def\bmx{\begin{matrix}}
    \def\emx{\end{matrix}}
\def\ben{\begin{enumerate}}
    \def\een{\end{enumerate}}
\def\bit{\begin{itemize}}
    \def\eit{\end{itemize}}
\def\bei{\begin{itemize}}
    \def\eei{\end{itemize}}
\def\bet{\begin{tabular}}
    \def\eet{\end{tabular}}

\newcommand{\ba}{\mathbf{a}}

\newcommand{\R}{\mathbb{R}}

\newcommand{\be}{\mathbf{e}}

\newcommand{\g}{\mathbf{g}}

\newcommand{\bu}{\mathbf{u}}

\newcommand{\bv}{\mathbf{v}}

\usepackage[dvipsnames]{xcolor}

\def\R{\mathbb{R}}

\def\1{{\bf1}}

\def\b{{\beta}}
\def\a{\alpha}
\def\g{\gamma}

\def\bit{\begin{itemize}}
\def\eit{\end{itemize}}

\def\be{\begin{equation}}
\def\ee{\end{equation}}
\def\ba{\begin{eqnarray}}
\def\ea{\end{eqnarray}}

\def\bes{\begin{equation*}}
\def\ees{\end{equation*}}
\def\bas{\begin{eqnarray*}}
\def\eas{\end{eqnarray*}}

\usepackage{url}
\usepackage{verbatim}

\newtheorem{Remark 1}{Remark}
\newtheorem{Remark 2}[Remark 1]{Remark}
\newtheorem{Remark 3}[Remark 1]{Remark}
\newtheorem{Remark 4}[Remark 1]{Remark}
\newtheorem{Remark 5}[Remark 1]{Remark}
\newtheorem{Remark 6}[Remark 1]{Remark}
\newtheorem{Remark 7}[Remark 1]{Remark}

\newtheorem{Lemma 1}{Lemma}
\newtheorem{Lemma 2}[Lemma 1]{Lemma}
\newtheorem{Lemma 3}[Lemma 1]{Lemma}
\newtheorem{Lemma 4}[Lemma 1]{Lemma}
\newtheorem{Lemma 5}[Lemma 1]{Lemma}
\newtheorem{Lemma 6}[Lemma 1]{Lemma}
\newtheorem{Lemma 7}[Lemma 1]{Lemma}

\newtheorem{Assumption 1}{Assumption}
\newtheorem{Assumption 2}[Assumption 1]{Assumption}
\newtheorem{Assumption 3}[Assumption 1]{Assumption}
\newtheorem{Assumption 4}[Assumption 1]{Assumption}
\newtheorem{Definition 1}{Definition}
\newtheorem{Theorem 1}{Theorem}
\newtheorem{Theorem 2}[Theorem 1]{Theorem}
\newtheorem{Theorem 3}[Theorem 1]{Theorem}
\newtheorem{Theorem 4}[Theorem 1]{Theorem}
\newtheorem{Theorem 5}[Theorem 1]{Theorem}
\newtheorem{Theorem 6}[Theorem 1]{Theorem}
\newtheorem{Theorem 7}[Theorem 1]{Theorem}
\newtheorem{Theorem 8}[Theorem 1]{Theorem}
\newtheorem{Theorem 9}[Theorem 1]{Theorem}
\newtheorem{Theorem 10}[Theorem 1]{Theorem}

\newtheorem{Proposition 1}{Proposition}

\title{\LARGE \bf
 Differentially-private Distributed Algorithms for Aggregative Games with Guaranteed Convergence}

\author{Yongqiang Wang, Angelia Nedi\'c
\thanks{   The work   was supported in part by the National Science Foundation under Grants   ECCS-1912702, CCF-2106293,   CCF-2106336, and CCF-2215088.}
\thanks{Yongqiang Wang is with the Department of Electrical and Computer Engineering, Clemson University, Clemson, SC 29634, USA
{\tt\small{yongqiw}@clemson.edu}
}%
\thanks{Angelia Nedi\'c is with the School of Electrical, Computer and Energy Engineering, Arizona State University, Tempe, AZ 85281, USA {\tt\small angelia.nedich@asu.edu}}
}

\begin{document}

\maketitle
\thispagestyle{empty}
\pagestyle{empty}

\begin{abstract}
The distributed computation of a Nash equilibrium in aggregative games is gaining increased traction in recent years. Of particular interest is the mediator-free scenario where individual players only access or observe the decisions of their neighbors  due to  practical constraints.  Given the competitive rivalry among  participating players, protecting the privacy of individual players becomes imperative when sensitive information is involved.   We propose a fully distributed equilibrium-computation approach for aggregative games that can achieve both rigorous differential privacy and guaranteed computation accuracy of the Nash equilibrium. This is in sharp contrast to existing differential-privacy solutions for aggregative games that have to either sacrifice the accuracy of equilibrium computation to gain rigorous privacy guarantees, or allow the cumulative privacy budget to grow unbounded, hence losing privacy guarantees, as  iteration proceeds. Our
approach uses independent noises across players, thus making
it effective   even when adversaries  have access to all shared messages as well as the underlying algorithm structure. The
encryption-free nature of the proposed approach,  also ensures efficiency in computation and communication. The approach is also applicable in stochastic aggregative games, able to ensure both rigorous differential privacy and guaranteed computation accuracy of the Nash equilibrium when individual players only have  stochastic estimates of their pseudo-gradient mappings.
 Numerical comparisons with existing counterparts confirm the effectiveness of the proposed approach.
\end{abstract}

\section{Introduction}
The   distributed computation of a Nash equilibrium over networks has gained increased attention in recent years. It has found applications  in various domains where multiple  players (agents)  compete to maximize their individual payoff functions, with typical examples including  energy management in smart grids \cite{saad2012game}, congestion control in communication networks \cite{yin2011nash}, market analysis in economics \cite{okuguchi2012theory}, and route coordination in road networks \cite{paccagnan2018nash}. In many of these application scenarios, a participating player's payoff function depends on the aggregate (e.g.,   total sum) of all  players' decisions, but such an aggregate is  inaccessible to individual players. Namely,   no central coordinator/mediator exits to collect and distribute the aggregate information, and a player can only access   the decisions of its immediate neighbors. Consequently, individual players cannot compute their accurate payoff functions, but instead, they share information among neighboring players to estimate the aggregate decision \cite{koshal2016distributed,salehisadaghiani2016distributed,pavel2019distributed,franci2021stochastic}.

Despite recent progress in such  aggregative games where distributed computation can be conducted under partial-decision information obtained through local information sharing   \cite{koshal2016distributed,salehisadaghiani2016distributed,pavel2019distributed,franci2021stochastic,nguyen2022distributed}, all these distributed algorithms explicitly share estimate/decision variables in every iteration, which can lead to a potential disclosure of players' sensitive information.  This problem is significant in that the players are opponents and have every reason to protect their individual private information in competition. Take the Nash-Cournot game as an example, players' cost functions  could be market sensitive and every player is well motivated to  protect its cost function to gain  an edge over its competitors \cite{rassenti2000adaptation}. Moreover, sometimes privacy legislations require  the privacy of players' information to be protected during  equilibrium-behavior implementation in a game. For example, in routing games \cite{dong2015differential},  California Privacy Rights Act forbids disclosing the  spatiotemporal information  of drivers  because  these information can be used as the basis for inferences of a person's activities \cite{gerhart2020proposition}.

To address the urgent need of privacy protection in aggregative games, some efforts have been reported in recent years (see, e.g., \cite{lu2015game,cummings2015privacy,shilov2021privacy}). However, these efforts mainly address Nash-equilibrium computation in the presence of a coordinator/mediator which  greatly simplifies the privacy design problem. The work \cite{gade2020privatizing} proposes  a privacy approach for fully distributed Nash-equilibrium computation, but the use of correlated noise restricts its applicability when   players can have arbitrary communication patterns. The paper \cite{shakarami2022distributed} proposes to use an uncertain parameter to obscure the pesudo-gradient mapping to enable privacy in continuous-time Nash-equilibrium computation algorithms. However, the fact that the  uncertain parameter  is a constant scalar restricts its privacy-protection strength. In fact, the approach can only avoid the payoff function from being uniquely identifiable, while the relations among  private parameters are still revealed. Given that differential privacy can provide strong protection against arbitrary post-processing and  auxiliary information \cite{dwork2014algorithmic}, and is becoming the de facto standard for privacy protection, the recent works  \cite{ye2021differentially} and \cite{wang2022differentially} propose  differential-privacy mechanisms for equilibrium computation in fully distributed aggregative games. However,  to ensure rigorous $\epsilon$-differential privacy (with finite cumulative privacy budget), these approaches have to sacrifice provable  convergence to the exact Nash equilibrium.

 To avoid the problem of trading  convergence accuracy for differential privacy  that is plaguing existing differential-privacy approaches for aggregative games,
    this paper presents the first distributed Nash-equilibrium computation approach that can simultaneously achieve both rigorous $\epsilon$-differential privacy (with finite cumulative privacy budget) and guaranteed convergence to the Nash equilibrium. Motivated by the observation that persistent differential-privacy noise has to be repeatedly injected in every iteration of information sharing, which results in significant reduction in algorithmic accuracy, our key idea is to gradually weaken the coupling strength  to attenuate the effect of differential-privacy noise added on shared messages.  We judiciously design the weakening factor sequence to ensure that convergence to the Nash equilibrium is guaranteed even in the presence of persistent differential-privacy noise. It is worth noting that compared with our recent result for differentially-private distributed optimization \cite{wang2022tailoring,wang2022quantization}, the results here are significantly different: 1) In  distributed optimization, agents cooperate to minimize a common objective function, whereas in aggregative games  players are competitive and only mind their own payoff functions; 2) Adding differential-privacy noise can easily alter the equilibrium of a game (just as evidenced by the loss of accurate convergence  in existing differential-privacy approaches for aggregative games \cite{ye2021differentially}), and hence we have to judiciously design our noise-adding mechanism to avoid perturbing the equilibrium; 3) Under the constraint of accurate convergence, our approach in \cite{wang2022tailoring} can only ensure bounded cumulative privacy budget of differential privacy in the vanilla single-variable gradient method,  and its cumulative privacy budget will still grow unbounded in the two-variable gradient-tracking based distributed optimization. In contrast, in this paper, we ensure a finite cumulative privacy budget for the proposed approach that involves   two iteration variables.

 {\bf Contributions:} The main contributions are summarized as follows:
\begin{enumerate}[wide, labelwidth=!,labelindent=8pt]
  \item   By judiciously designing the aggregate estimation mechanism, we propose a fully distributed computation approach for aggregative games  that can ensure rigorous $\epsilon$-differential privacy without losing guaranteed convergence to the Nash equilibrium. The algorithm can ensure both a finite cumulative privacy budget and accurate convergence, which is in sharp contrast to existing differential-privacy approaches for aggregative games (see, e.g., \cite{ye2021differentially} and \cite{wang2022differentially}) that have to trade accurate convergence for differential privacy.   To the best of our knowledge, this is the first such algorithm in the literature.
  \vspace{0.15cm}
  \item  We propose a new proof technique for the convergence analysis of the fully distributed computation approach for aggregative games in the presence of information-sharing noise (caused by, e.g., differential-privacy design). The new convergence derivation does not impose the restriction that the pseudo-gradient mapping is uniformly bounded, an assumption  that is used in existing distributed algorithms (e.g., \cite{ye2021differentially} and \cite{wang2022differentially}) for aggregative games subject to noises. Note that avoiding the uniformly bounded pseudo-gradient assumption is significant since in the presence of differential-privacy noise (e.g., Laplacian or Gaussian noise) which are not uniformly bounded, the  aggregative estimation may become unbounded, which makes the pseudo-gradient mapping   unbounded in many common games such as the Nash-Cournot game under a price governed by the linear inverse-demand function.
   \vspace{0.15cm}
  \item  Even without taking privacy into consideration, the  proposed algorithms  and theoretical derivations  are of interest themselves.
  The convergence analysis for the   proposed algorithms  has fundamental differences from existing proof techniques. More specifically, existing convergence analysis of distributed (generalized) Nash-equilibrium computation algorithms for aggregative games (e.g., \cite{koshal2016distributed,belgioioso2020distributed,parise2019distributed,gadjov2020single,zhu2022asynchronous,liang2017distributed}) and their stochastic variants (e.g., \cite{wang2022differentially} and \cite{lei2022distributed}) rely on the geometric (exponential) decreasing of the  aggregate-estimation error (consensus error) among the players, which is possible only when all nonzero coupling weights are lower bounded by a positive constant. Such geometric (exponential) decreasing of aggregate-estimation error is key to proving exact convergence of all players' iterates to the Nash equilibrium. In our case, since the coupling strength decays to zero, such geometric (exponential)  decreasing of players' aggregate-estimation error does not exist any more, which makes it impossible to use the proof techniques in existing results.

   \vspace{0.15cm}
  \item  We extend  the approach to the case where the pseudo-gradient mapping is stochastic, and prove that rigorous $\epsilon$-differential privacy and guaranteed convergence can still be achieved simultaneously in this case.  Note that different from \cite{wang2022differentially,lei2022distributed} which consider  stochastic pseudo-gradients with decreasing variances (via increasing sample sizes), we allow the variance of the stochastic pseudo-gradient  to be constant, or even increasing with time, as specified in Remark \ref{re:variance_gradients}.
\end{enumerate}

The organization of the paper is as follows.
Sec.~\ref{sec-problem} gives the problem formulation
and some results for a later use. Sec. \ref{se:algorithm1} presents a differentially-private distributed computation algorithm for aggregative games. This section also proves that  the algorithm can ensure all players'  convergence to the exact Nash equilibrium while ensuring rigorous $\epsilon$-differential privacy with a finite cumulative privacy budget, even when the number of iterations goes to infinity. Sec. \ref{se:algorithm2} extends the approach to the case of stochastic aggregative games and prove that it can ensure both guaranteed computation accuracy of the Nash equilibrium and  differential privacy with guaranteed finite cumulative privacy budget  when individual players only have  stochastic estimates of their pseudo-gradient mappings. Sec. \ref{se:simulation} presents numerical comparisons with existing distributed computation approaches for aggregative games to confirm the obtained results. Finally, Sec. \ref{se:conclusions} concludes the paper.

{\bf Notations:}  We use $\mathbb{R}^d$ to denote the Euclidean space of
dimension $d$. We write $I_d$ for the identity matrix of dimension $d$,
and ${\bf 1}_d$ for  the $d$-dimensional  column vector will all
entries equal to 1; in both cases  we suppress the dimension when it is
clear from the context. 
A vector is viewed as a column
vector, and for  a
vector $x$, $[x]_i$ denotes its $i$th element.
We write $x> 0$ (resp. $x\geq 0$) if all elements of
$x$ are positive (resp. non-negative).
  We use $\langle\cdot,\cdot\rangle$ to denote the inner product and
 $\|x\|$ for the standard Euclidean norm of a vector $x$. We use $\|x\|_1$ to represent the $L_1$ norm of a vector $x$.
We write $\|A\|$ for the matrix norm induced by the vector norm $\|\cdot\|$.
We let $A^T$ denote the transpose of a matrix $A$.
 For two  vectors  $u$ and $v$ with the same dimension, we use $u\leq v$ to represent the relationship $u-v\leq 0$. Often, we abbreviate {\it almost surely} by {\it a.s}.
\def\as{{\it a.s.\ }}


\section{Problem Formulation and Preliminaries}\label{sec-problem}

\subsection{On Aggregative Games}
We consider  a set of $m$ players (or agents), i.e., $[m]=\{1,2,\ldots,m\}$, which are indexed by $1,2,\cdots,m$. Player $i$ is characterized by a strategy set $K_i \subseteq\mathbb{R}^d$ and a payoff function $f_i(x_i,\bar{x})$ where $x_i$ denotes the decision of player $i$ and $\bar{x}=\frac{1}{m}\sum_{i=1}^{m}x_i$  denotes the average of all  players' decisions. Note that sometimes the payoff function depends on the aggregate decision $m\bar{x}=\sum_{i=1}^{m}x_i$ rather than the average decision $\bar{x}$. In these cases, all algorithms and analysis in this paper are still valid by replacing $\bar{x}$ with $m\bar{x}$.

 Since each decision variable $x_i$ is restricted in $K_i$, the average $\bar{x}$ is restricted to the set
that is the $1/m$-scaling of the Minkowski sum\footnote{A scaling $tX$ of a set $X$ with a scalar $t$ is the set given by
$tX=\{tx\mid x\in X\}$.
A Minkowski sum of two sets $X$ and $Y$ is the set
$X+Y=\{x+y\mid x\in X, y\in Y\}$.} of the sets $K_i$, denoted by $\bar K$,
i.e.,
\[\bar K=\frac{1}{m}(K_1+K_2+\cdots+K_m).\]
With this notation, we can formalize $f_i(x_i,\bar{x})$   as a  mapping from $K_i\times\bar{K}$ to $\mathbb{R}$, and further formulate the game that player $i$ faces as the following parameterized optimization problem:
\begin{equation}\label{eq:formulation}
 \min f_i(x_i,\bar{x})\quad {\rm s.t.}\quad x_i\in K_i\:\: {\rm and}\:\: \bar{x}\in\bar{K}.
\end{equation}
The constraint set $K_i$ and the function $f_i(\cdot)$ are assumed to be known to player $i$ only.

To characterize  a Nash equilibrium of the aggregative game (\ref{eq:formulation}), following \cite{koshal2016distributed}, we introduce the following notations
\begin{equation}\label{eq:F_i}
 F_i(x_i,\bar{x})\triangleq \nabla_{x_i} f_i(x_i,\bar{x}),
\end{equation}
\begin{equation}\label{eq:K}
K\triangleq \Pi_{i=1}^m K_i,
\end{equation}
and
\[
x\triangleq \left[x_1^T,\cdots,x_m^T\right]^T.
\]

These notions allow us to define two mappings
\begin{equation}\label{eq:F}
 F(x,u)\triangleq \left(\begin{array}{c}F_1(x_1,u)\\ \vdots\\F_m(x_m,u)\end{array}\right),
\end{equation}
\begin{equation}
\phi(x)=F(x,\bar{x}),\quad \forall x\in K.
\end{equation}

Similar to \cite{koshal2016distributed}, we make the following assumptions on the constraint sets $K_i$ and the functions $f_i$:
\begin{Assumption 2}\label{as:functions}
Each $K_i\in\mathbb{R}^d$ is compact and convex. Each function $f_i(x_i,y)$ is continuously differentiable in $(x_i,y)$ over some open set containing the set $K_i\times \bar{K}$, while each function $f_i(x_i,\bar{x})$ is convex in $x_i$ over the set $K_i$. The mapping $\phi(x)$ is strictly monotone over $K$, i.e., for all $x\neq x'$ in $K$, we always have
\[
\left(\phi(x)-\phi(x')\right)^T(x-x')>0.
\]
\end{Assumption 2}
\begin{Remark 1}
  It is worth noting that the strictly monotone assumption on $\phi(x)$ is weaker than the commonly used strongly monotone assumption  in \cite{pavel2019distributed,ye2021differentially,wang2022differentially,lei2022distributed,yousefian2015self,tatarenko2020geometric}.
\end{Remark 1}

According to \cite{koshal2016distributed}, Assumption \ref{as:functions} ensures that the aggregative game (\ref{eq:formulation}) has a unique Nash equilibrium   $ x^{\ast}=
[(x_1^{\ast})^T,(x_2^{\ast})^T,\ldots,(x_m^{\ast})^T]^T\in\mathbb{R}^{md}$. Moreover, following \cite{koshal2016distributed}, we also make the following assumption on the  mapping $F_i(x_i,u)$:
\begin{Assumption 2}\label{as:Lipschitz}
  Each mapping $F_i(x_i,u)$ satisfies the following Lipschitz continuous condition with respect to $u$:   for all
$x_i\in K_i$, all $u_1,u_2\in\bar{K}$, and all $i\in[m]$,  we always have
\[
\|F_i(x_i,u_1)-F_i(x_i,u_2) \|\leq \tilde{L}\|u_1-u_2\|
\]
for some $\tilde{L}>0$.
\end{Assumption 2}

We consider distributed algorithms for equilibrium computation of the game in (\ref{eq:formulation}). Namely, no player has a direct access to the average decision $\bar{x}$ or the aggregate decision $m\bar{x}$. Instead, each player has to construct a local estimate of the average/aggregate through local interactions with its neighbors. We describe the local interaction using a weight matrix
$L=\{L_{ij}\}$, where $L_{ij}>0$ if   player  $j$ and player $i$ can directly communicate with each other,
and $L_{ij}=0$ otherwise. For a player $i\in[m]$,
its  neighbor set
$\mathbb{N}_i$ is defined as the collection of players $j$ such that $L_{ij}>0$.
We define $L_{ii}\triangleq-\sum_{j\in\mathbb{N}_i}L_{ij}$  for all $i\in [m]$,
where $\mathbb{N}_i$ is the neighbor set of agent $i$. Furthermore,
We make the following assumption on $L$:

\begin{Assumption 1}\label{as:L}
 The matrix  $L=\{w_{ij}\}\in \mathbb{R}^{m\times m}$ is symmetric and satisfies
    ${\bf 1}^TL={\bf
  0}^T$, $L{\bf 1}={\bf
  0}$, and $ \|I+L-\frac{{\bf 1}{\bf 1}^T}{m}\|<1$.
\end{Assumption 1}

Assumption~\ref{as:L} ensures that the interaction graph induced by $L$ is connected, i.e., there is a  path
from each player to every other player.
It can be verified that $\|I+L-\frac{{\bf 1}{\bf 1}^T}{m}\|=\max\{|1+\rho_2|,|1+\rho_m|\}$, where $\{\rho_i,i\in[m]\}$ are the eigenvalues of $L$, with
$\rho_m\le \ldots\le \rho_2\le\rho_1=0$.

 In the analysis of our methods, we use the following results:
\begin{Lemma 2}\cite{wang2022tailoring}\label{Lemma-polyak_2}
Let $\{v^k\}$,$\{\a^k\}$, and $\{p^k\}$ be random nonnegative scalar sequences, and
$\{q^k\}$ be a deterministic nonnegative scalar sequence satisfying
$\sum_{k=0}^\infty \a^k<\infty$  almost surely,
$\sum_{k=0}^\infty q^k=\infty$, $\sum_{k=0}^\infty p^k<\infty$ almost surely,
and the following inequality:
\[
\mathbb{E}\left[v^{k+1}|\mathcal{F}^k\right]\le(1+\a^k-q^k) v^k +p^k,\quad \forall k\geq 0\quad\as
\]
where $\mathcal{F}^k=\{v^\ell,\a^\ell,p^\ell; 0\le \ell\le k\}$.
Then, $\sum_{k=0}^{\infty}q^k v^k<\infty$ and
$\lim_{k\to\infty} v^k=0$ hold almost surely.
\end{Lemma 2}
\begin{Lemma 1}\cite{wang2022tailoring}\label{le:vector_convergence}
Let  $\{\bv^k\}\subset \mathbb{R}^d$
and $\{\bu^k\}\subset \mathbb{R}^p$ be random nonnegative
vector sequences, and $\{a^k\}$ and $\{b^k\}$ be random nonnegative scalar sequences   such that
\[
\mathbb{E}\left[\bv^{k+1}|\mathcal{F}^k\right]\le (V^k+a^k{\bf 1}{\bf1}^T)\bv^k +b^k{\bf 1} -H^k\bu^k,\quad \forall k\geq 0
\]
holds almost surely, where $\{V^k\}$ and $\{H^k\}$ are random sequences of
nonnegative matrices and
$\mathbb{E}\left[\bv^{k+1}|\mathcal{F}^k \right]$ denotes the conditional expectation given
 $\bv^\ell,\bu^\ell,a^\ell,b^\ell,V^\ell,H^\ell$ for $\ell=0,1,\ldots,k$.
Assume that $\{a^k\}$ and $\{b^k\}$ satisfy
$\sum_{k=0}^\infty a^k<\infty$ and $\sum_{k=0}^\infty b^k<\infty$ almost surely, and
that there exists a (deterministic) vector $\pi>0$ such that
$\pi^T V^k\le \pi^T$ and $\pi^TH^k\ge 0$ hold almost surely for all $k\geq 0$.
Then, we have
1) $\{\pi^T\bv^k\}$ converges to some random variable $\pi^T\bv\geq 0$ almost surely; 2) $\{\bv^k\}$ is bounded almost surely; and
3) $\sum_{ k=0 }^\infty \pi^TH^k\bu^k<\infty$ holds almost surely.
\end{Lemma 1}

\subsection{On Differential Privacy}
We adopt the notion of $\epsilon$-differential privacy for continuous bit streams \cite{dwork2010differential}, which has recently been applied to distributed optimization algorithms (see \cite{Huang15} as well as our work \cite{wang2022tailoring}). A commonly used approach to enable differential privacy is injecting Laplace   noise to shared messages. For a constant $\nu>0$, we use ${\rm Lap}(\nu)$ to denote a Laplace distribution of a scalar random variable with the probability density function $x\mapsto\frac{1}{2\nu}e^{-\frac{|x|}{\nu}}$. It can be verified that ${\rm Lap}(\nu)$ has  zero mean  and variance $2\nu^2$.
Following the formulation of distributed optimization  in \cite{Huang15}, for the convenience of differential-privacy analysis, we represent the distributed game $\mathcal{P}$ in (\ref{eq:formulation}) by three parameters ($K, \mathbb{F}, L$), where
 $K$ defined in (\ref{eq:K}) is the domain of decision variables,     $ \mathbb{F} \triangleq\{ f_1,\,\cdots,f_m\}$, and $ L$ is the inter-player interaction weight matrix $L$. Then we define adjacency between two games as follows:

\begin{Definition 1}\label{de:adjacency}
Two distributed Nash-equilibrium computation problems $\mathcal{P}=( K, \mathbb{F}, L)$ and $\mathcal{P}'=(K', \mathbb{F}', L')$ are adjacent if the following conditions hold:
\begin{itemize}
\item $K=K'$  and $L=L'$, i.e., the domains of decision variables   and the interaction weight matrices are identical;
\item there exists an $i\in[m]$ such that $f_i\neq f_i'$ but $f_j=f_j'$ for all $j\in[m],\,j\neq i$.
\end{itemize}
\end{Definition 1}

According to Definition \ref{de:adjacency}, it can be seen that two distributed aggregative games are adjacent if and only if one player changes its payoff function (can be in an arbitrary way) while all other game characteristics are identical.

 Given a distributed Nash-equilibrium computation algorithm, we represent an execution of such an algorithm as $\mathcal{A}$, which is an infinite sequence of the iteration variable $\vartheta$, i.e., $\mathcal{A}=\{\vartheta^0,\vartheta^1,\cdots\}$. We consider adversaries that can observe all communicated messages in the network. Therefore, the observation part of an execution is the infinite sequence of shared messages, which is represented by $\mathcal{O}$. We define the observation mapping as $\mathcal{R}(\mathcal{A})\triangleq \mathcal{O}$. Given a distributed Nash-equilibrium computation problem $\mathcal{P}$, observation sequence $\mathcal{O}$, and an initial state $\vartheta^0$,  $\mathcal{R}^{-1}(\mathcal{P},\mathcal{O},\vartheta^0)$ is the set of executions $\mathcal{A}$ that can generate the observation $\mathcal{O}$.
 \begin{Definition 1}
   ($\epsilon$-differential privacy, adapted from \cite{Huang15}). For a given $\epsilon>0$, an iterative distributed algorithm solving problem~(\ref{eq:formulation}) is $\epsilon$-differentially private if for any two adjacent $\mathcal{P}$ and $\mathcal{P}'$, any set of observation sequences $\mathcal{O}_s\subseteq\mathbb{O}$ (with $\mathbb{O}$ denoting the set of all possible observation sequences), and any initial state ${\vartheta}^0$, we always have
    \begin{equation}
        \mathbb{P}[\mathcal{R}^{-1}\left(\mathcal{P},\mathcal{O}_s,{\vartheta}^0\right)]\leq e^\epsilon\mathbb{P}[\mathcal{R}^{-1}\left(\mathcal{P}',\mathcal{O}_s,{\vartheta}^0\right)],
    \end{equation}
    where the probability $\mathbb{P}$ is taken over the randomness over iteration processes.
 \end{Definition 1}

The above definition of $\epsilon$-differential privacy ensures that an adversary having access to all shared messages in the network cannot gain information with a  significant probability of any participating player's payoff function. It can also be seen that a smaller $\epsilon$ means a higher level of privacy protection. It is also worth noting that the considered notion of $\epsilon$-differential privacy is more stringent than other relaxed (approximate) differential privacy notions such as $(\epsilon,\,\delta)$-differential privacy \cite{kairouz2015composition}, zero-concentrated differential privacy \cite{bun2016concentrated}, or R\'{e}nyi differential privacy \cite{mironov2017renyi}.

\section{A differentially-private distributed computation algorithm for aggregative games}\label{se:algorithm1}

To achieve strong differential privacy, independent noise should be injected  repeatedly in every round of message sharing and, hence, constantly affecting the algorithm through inter-player interactions and leading to significant reduction in algorithmic accuracy.
Motivated by this observation, we propose to gradually weaken inter-player interactions to reduce the influence of differential-privacy noise on computation accuracy. Interestingly, we prove that by judiciously designing the interaction weakening mechanism, we can  ensure  convergence of all players   to the Nash equilibrium  even in the presence of persistent
differential-privacy noise.

\noindent\rule{0.49\textwidth}{0.5pt}
\noindent\textbf{Algorithm 1: Differentially-private distributed algorithm for aggregative games with guaranteed convergence}

\noindent\rule{0.49\textwidth}{0.5pt}
\begin{enumerate}[wide, labelwidth=!, labelindent=0pt]
    \item[] Parameters: Stepsize $\lambda^k>0$ and
    weakening factor $\gamma^k>0$.
    \item[] Every player $i$ maintains one decision variable  $x_i^k$, which is initialized with a random vector in $K_i\subseteq\mathbb{R}^d$, and an estimate of the aggregate decision $v_i^k$, which is initialized as $v_i^0=x_i^0$.
    \item[] {\bf for  $k=1,2,\ldots$ do}
    \begin{enumerate}
        \item Every player $j$ adds persistent differential-privacy noise   $\zeta_j^{k}$ 
        to its estimate
    $v_j^k$,  and then sends the obscured estimate $v_j^k+\zeta_j^{k}$ to player
        $i\in\mathbb{N}_j$.
        \item After receiving  $v_j^k+\zeta_j^k$ from all $j\in\mathbb{N}_i$, player $i$ updates its decision variable and estimate  as follows:
        \begin{equation}\label{eq:update_in_Algorithm1}
        \begin{aligned}
             x_i^{k+1}&=\Pi_{K_i}\left[x_i^k-\lambda^k F_i(x_i^k,v_i^k)\right],\\
             v_i^{k+1}&=v_i^k+\gamma^k\sum_{j\in \mathbb{N}_i} L_{ij}(v_j^k+\zeta_j^k-v_i^k-\zeta_i^k)+x_i^{k+1}-x_i^k,
        \end{aligned}
        \end{equation}
        where  $\Pi_{K_i}[\cdot]$ denotes the Euclidean projection of a vector onto the set $K_i$.
                \item {\bf end}
    \end{enumerate}
\end{enumerate}
\vspace{-0.1cm} \rule{0.49\textwidth}{0.5pt}
\begin{Remark 1}
 In the iterates in (\ref{eq:update_in_Algorithm1}), we judiciously let player $i$ use $v_i^k+\zeta_i^k$ that it shares with its neighbors in its interaction terms ($L_{ij}(v_j^k+\zeta_j^k-v_i^k-\zeta_i^k)$ for $j\in \mathbb{N}_i$) to cancel out the influence of noises on the aggregate estimation (average estimation, more precisely). As shown latter in Lemma \ref{le:bar_x=bar_v}, player $i$ using $v_i^k+\zeta_i^k$ in its interaction terms rather than $v_i^k$ is key to ensure that the average $v_i^k$ among all players can accurately track the average decision $x_i^k$ among all players.  Note that different from \cite{gade2020privatizing} where players use correlated noise which restricts the strength of privacy protection, here the noises $\zeta_i^k$ ($i=1,\,2,\,\cdots,m$) of all players are completely independent of each other, and hence can enable strong differential privacy.
\end{Remark 1}

 The sequence $\{\gamma^k\}$, which diminishes with time, is used to suppress the influence of persistent differential-privacy noise $\zeta_j^k$ on the convergence point of the iterates.
The stepsize sequence $\{\lambda^k\}$ and attenuation sequence $\{\gamma^k\}$
have to be designed appropriately to guarantee the accurate convergence of the iterate vector $x^k\triangleq[(x_1^k)^T,\cdots,(x_m^k)^T]^T$ to the Nash equilibrium point $x^{\ast}\triangleq[(x_1^\ast)^T,\cdots,(x_m^\ast)^T]^T$.
The persistent differential-privacy noise sequences $\{\zeta_i^k\}, i\in[m]$ have zero-mean and
$\gamma^k$-bounded  (conditional) variances, which will be specified later in Assumption \ref{assumption:dp-noise}.

\subsection{Convergence Analysis}

To prove the convergence of  the decision vector $x^k$   to the Nash equilibrium $x^{\ast}$, we have to present some properties of the iterates. The first property pertains to the average of the estimates $v_i^k$, which is defined as $ \bar{v}^k\triangleq\frac{1}{m}\sum_{i=1}^{m} v_i^k$.
More specifically, we will prove that $\bar{v}^k$ is equal to the average of decisions $\bar{x}^k \triangleq\frac{1}{m}\sum_{i=1}^{m} x_i^k$.
Namely, $\bar{v}^k$ captures the exact average decision. Such a property has been proven and employed in \cite{koshal2016distributed} in the absence of noise. Now we prove that this relationship still holds under our proposed Algorithm 1 even all agents add independent noises to their shared messages.

\begin{Lemma 1}\label{le:bar_x=bar_v}
Under Assumption \ref{as:L}, we have $\bar{v}^k=\bar{x}^k$ for all $k\geq 0$.
\end{Lemma 1}
\begin{proof}

According to the definitions of $\bar{v}^k$ and $\bar{x}^k$, we only have to prove
\begin{equation}\label{eq:conservation_average}
\sum_{i=1}^{m}v_i^k=\sum_{i=1}^{m}x_i^k.
\end{equation}
 We prove the relationship in (\ref{eq:conservation_average}) using induction.

  For $k=0$, the relationship holds trivially since we have initialized all $v_i^k$ as $v_i^0=x_i^0$.

 Next we proceed to prove that if (\ref{eq:conservation_average}) holds for some iteration $k>0$, i.e.,
  \begin{equation}\label{eq:equation_at_k1}
 \sum_{i=1}^{m} v_i^k=\sum_{i=1}^{m} x_i^k,
 \end{equation}
  then it also holds for iteration $k+1$.

 According to (\ref{eq:update_in_Algorithm1}), we have
 \begin{equation}\label{eq:iteration_in_lemma}
 \begin{aligned}
 \sum_{i=1}^{m}v_i^{k+1}=&\sum_{i=1}^{m} v_i^k+\gamma^k\sum_{i=1}^{m} \sum_{j\in \mathbb{N}_i} L_{ij}(v_j^k+\zeta_j^k-v_i^k-\zeta_i^k)\\
            &+\sum_{i=1}^{m} x_i^{k+1}-\sum_{i=1}^{m} x_i^k.
 \end{aligned}
 \end{equation}

 Plugging (\ref{eq:equation_at_k1}) into (\ref{eq:iteration_in_lemma}) leads to
  \begin{equation}\label{eq:iteration_in_lemma2}
 \begin{aligned}
 \sum_{i=1}^{m}v_i^{k+1}=&\gamma^k\sum_{i=1}^{m} \sum_{j\in \mathbb{N}_i} L_{ij}(v_j^k+\zeta_j^k-v_i^k-\zeta_i^k)+\sum_{i=1}^{m} x_i^{k+1}.
 \end{aligned}
 \end{equation}

 We decompose the first term (excluding $\gamma^k$) on the right hand side of (\ref{eq:iteration_in_lemma2}) as
  \begin{equation}
 \begin{aligned}
  &\sum_{i=1}^{m} \sum_{j\in \mathbb{N}_i} L_{ij}(v_j^k+\zeta_j^k-v_i^k-\zeta_i^k)\\
  &= \sum_{i=1}^{m} \sum_{j\in \mathbb{N}_i} L_{ij}v_j^k-\sum_{i=1}^{m} \sum_{j\in \mathbb{N}_i} L_{ij}v_i^k +\sum_{i=1}^{m} \sum_{j\in \mathbb{N}_i} L_{ij} \zeta_j^k\\
            &\qquad-\sum_{i=1}^{m} \sum_{j\in \mathbb{N}_i} L_{ij} \zeta_i^k.
  \end{aligned}
 \end{equation}
 Using the symmetric property of $L_{ij}$ in Assumption \ref{as:L}, the preceding relationship can be rewritten as
 \begin{equation}\label{eq:symmetric}
 \begin{aligned}
  &\sum_{i=1}^{m} \sum_{j\in \mathbb{N}_i} L_{ij}(v_j^k+\zeta_j^k-v_i^k-\zeta_i^k)\\
  &= \sum_{i=1}^{m} \sum_{j\in \mathbb{N}_i} L_{ij}v_j^k-\sum_{i=1}^{m} \sum_{i\in \mathbb{N}_j} L_{ji}v_i^k +\sum_{i=1}^{m} \sum_{j\in \mathbb{N}_i} L_{ij} \zeta_j^k\\
            &\qquad-\sum_{i=1}^{m} \sum_{i\in \mathbb{N}_j} L_{ji} \zeta_i^k \\
  &=0.
  \end{aligned}
 \end{equation}
 Plugging (\ref{eq:symmetric}) into (\ref{eq:iteration_in_lemma2}) leads to
$\sum_{i=1}^{m} v_i^{k+1}=\sum_{i=1}^{m} x_i^{k+1}$,
 which completes the proof.
 \end{proof}

Using Lemma \ref{le:bar_x=bar_v}, we have the following results under Assumption \ref{as:functions}:
\begin{Lemma 1}\label{le:bounded}
 Under Assumption \ref{as:functions}, and $v_i^k$ governed by Algorithm 1, the following inequalities hold for some $C>0$ and all $k\geq 0$:
 \begin{equation}\label{eq:bounded_gradient}
  \|F_i(x_i^k,\bar{v}^k)\|\leq C,\quad \|F_i(x_i^k,v_i^k)\|\leq C+\tilde{L}\|v_i^k-\bar{v}^k\|.
 \end{equation}
\end{Lemma 1}
\begin{proof}
According to Lemma \ref{le:bar_x=bar_v}, we have $\bar{v}^k=\bar{x}^k$ for all $k\geq 0$, where $\bar{x}^k\triangleq \frac{\sum_{i=1}^{m}x_i^k}{m}$. Hence, $\bar{v}^k\in\bar{K}$, where $\bar{K}$ is compact since each $K_i$ is compact according to Assumption \ref{as:functions}. From Assumption \ref{as:functions}, $F_i(x_i^k,\bar{x}^k)$ is continuous over $K_i\times \bar{K}$, so we have the first inequality.

To show the second inequality, we use the following relationship
\[
\begin{aligned}
\|F_i(x_i^k,v_i^k)\|&=\|F_i(x_i^k,v_i^k)-F_i(x_i^k,\bar{x}^k)+F_i(x_i^k,\bar{x}^k)\|\\
&\leq \|F_i(x_i^k,v_i^k)-F_i(x_i^k,\bar{v}^k)\|+\|F_i(x_i^k,\bar{v}^k)\|.
\end{aligned}
\]
Then using the Lipschitz continuous condition in Assumption \ref{as:Lipschitz} and the proven fact that $\|F_i(x_i^k,\bar{v}^k)\|$ is bounded, we can arrive at the second inequality in (\ref{eq:bounded_gradient}).
\end{proof}
\begin{Remark 1}\label{re:boundedness}
   Note that different from \cite{ye2021differentially,wang2022differentially} whose convergence analysis requires $F_i(x_i,v_i^k)$ to be uniformly bounded in the presence of noise, we will provide a new proof technique that  removes the uniformly bounded  constraint in convergence analysis. This relaxation is significant in that under differential-privacy design, $v_i^k$ will be subject to unbounded noise, such as Laplace noise or Gaussian noise, and becomes unbounded. Therefore,  restricting $F_i(x_i,v_i^k)$ to be uniformly bounded with respect to $v_i^k$ will significantly limit the applicability of the algorithm. For example, in the  Nash-Cournot market game considered in the numerical simulations in Sec. \ref{se:simulation}, the sale price function (the inverse demand function) is usually modeled as  a function decreasing linearly with the aggregative production, which will result in a mapping $F_i(x_i,v_i^k)$ that is not uniformly bounded.
\end{Remark 1}

We   now apply Lemma~\ref{le:vector_convergence} to arrive at a general convergence theory for  distributed algorithms for the problem in (\ref{eq:formulation}):

\begin{Proposition 1}\label{th-main_decreasing}
Assume that problem (1) has a Nash equilibrium   $ x^{\ast}=
[(x_1^{\ast})^T,(x_2^{\ast})^T,\ldots,(x_m^{\ast})^T]^T\in\mathbb{R}^{md}$. Suppose that a distributed algorithm generates sequences
$\{x_i^k\}\subseteq\R^d$ and $\{v_i^k\}\subseteq\R^d$ such that
almost surely we have
\begin{equation}\label{eq:Theorem_decreasing}
\begin{aligned}
&\left[\begin{array}{c}
\mathbb{E}\left[\sum_{i=1}^m\|x_i^{k+1}-x_i^*\|^2|\mathcal{F}^k\right]\cr
\mathbb{E}\left[\sum_{i=1}^m\|v_i^{k+1}-\bar{v}^{k+1}\|^2|\mathcal{F}^k\right]\end{array}
\right]
\\
&\le \left( \left[\begin{array}{cc}
1 &  \kappa_1\gamma^k \cr
0& 1-\kappa_2\gamma^k\cr
\end{array}\right]
+a^k {\bf 1}{\bf 1}^T\right)\left[\begin{array}{c}\sum_{i=1}^m\|x_i^{k}-x_i^*\|^2\cr
\sum_{i=1}^m\|v_i^k-\bar{v}^k\|^2\end{array}\right]&&\cr
&\quad+b^k{\bf 1} - c^k \left[\begin{array}{c}
 \left(\phi(x^k)- \phi(x^*)\right)^T(x^k-x^{\ast}) \cr
 0\end{array}\right],\quad\forall k\geq 0
 \end{aligned}
\end{equation}
where  $\bar{v}^k=\frac{1}{m}\sum_{i=1}^m v_i^k$,
$\mathcal{F}^k=\{x_i^\ell,\,v_i^\ell, \, i\in[m],\, 0\le \ell\le k\}$,
the random
nonnegative scalar sequences $\{a^k\}$, $\{b^k\}$ satisfy $\sum_{k=0}^\infty a^k<\infty$ and $\sum_{k=0}^\infty b^k<\infty$ almost surely,  the deterministic  nonnegative sequences $\{c^k\}$ and $\{\gamma^k\}$  satisfy
$
\sum_{k=0}^\infty c^k=\infty$ and $
\sum_{k=0}^\infty \gamma^k=\infty
$, and  the scalars $\kappa_1$ and $\kappa_2$ satisfy  $\kappa_1>0$ and $0<\kappa_2\gamma^k<1$, respectively, for all $k\geq 0$.
Then, we have
$\lim_{k\to\infty}\|v_i^k - \bar v^k\|=0$ almost surely for all   $i$,
and
$\lim_{k\to\infty}\|x_i^k-x_i^*\|=0$ almost surely.
\end{Proposition 1}
\begin{proof}
According to Assumption \ref{as:functions}, we always have
 $\left(\phi(x^k)- \phi(x^*)\right)^T(x^k-x^{\ast})>0$ for all $k$.
Hence,  by letting $\bv^k=\left[\sum_{i=1}^{m}\|x_i^k-x_i^*\|^2,\ \sum_{i=1}^m \|v_i^k-\bar v^k\|^2\right]^T$,
from relation~(\ref{eq:Theorem_decreasing}) it follows that almost surely for all $k\ge0$,
\be\label{eq-fin0}
\hspace{-0.2cm}\mathbb{E}\left[\bv^{k+1}|\mathcal{F}^k\right]
\le \left( \left[\begin{array}{cc}
1 &  \kappa_1\gamma^k \cr
0& 1-\kappa_2\gamma^k\cr\end{array}\right] +a^k {\bf 1}{\bf 1}^T \right)\bv^k+b^k{\bf 1}.
\ee
Consider the vector $\pi=[1, \frac{\kappa_1}{\kappa_2}]^T$ and note
\[
\pi^T \left[\begin{array}{cc}
1 &  \kappa_1\gamma^k  \cr
0& 1-\kappa_2\gamma^k\cr\end{array}\right]= \pi^T.
\]
Thus, relation~\eqref{eq-fin0} satisfies all   conditions of Lemma~\ref{le:vector_convergence}.
By Lemma~\ref{le:vector_convergence}, it follows that
  $\lim_{k\to\infty}\pi^T\bv^k$ exists almost surely, and that the sequences $\{\sum_{i=1}^{m}\|x_i^k-x_i^*\|^2\}$
and $\{\sum_{i=1}^m \|v_i^k-\bar v^k\|^2\}$ are bounded almost surely.
From \eqref{eq-fin0}, we have the following relation almost surely  for the second element of $\bv^k$:
\begin{equation*}
\begin{aligned}
&\mathbb{E}\left[\sum_{i=1}^m \|v_i^{k+1}-\bar v^{k+1}\|^2|\mathcal{F}^k\right]\cr
&\le  (1+a^k -\kappa_2\gamma^k)\sum_{i=1}^m \|v_i^k - \bar v^k\|^2 + \b^k\quad\forall k\ge0,
\end{aligned}
\end{equation*}
where
$\b^k=a^k\left( \sum_{i=1}^m\left(\|x_i^k - x_i^\ast\|^2+ \|v_i^k - \bar v^k\|^2\right)\right)$.
 Since
$\sum_{k=0}^\infty a^k<\infty$ holds almost surely by our assumption, and
the sequences $\{\sum_{i=1}^m\|x_i^k-x_i^*\|^2\}$
and $\{\sum_{i=1}^m \|v_i^k-\bar v^k\|^2\}$ are bounded almost surely, it follows that $\sum_{k=0}^\infty\b^k<\infty$ holds almost surely.
Thus, the preceding relation satisfies the conditions of
Lemma~\ref{Lemma-polyak_2} with
$v^k= \sum_{i=1}^m \|v_i^k - \bar v^k\|^2$, $q^k=\kappa_2\gamma^k$,  and $p^k=\b^k$
due to our assumptions $\sum_{k=0}^\infty b^k<\infty$ almost surely and
$\sum_{k=0}^\infty \g^k=\infty$.
By Lemma \ref{Lemma-polyak_2}, it follows that almost surely
\be\label{eq-sumable}
 \sum_{k=0}^\infty \kappa_2\gamma^k\sum_{i=1}^m \|v_i^k - \bar v^k\|^2<\infty,\:
 \lim_{k\to\infty} \sum_{i=1}^m \|v_i^k - \bar v^k\|^2=0.
 \ee

 It remains to show that $\sum_{i=1}^{m}\|x_i^k-x_i^*\|^2\to0$ almost surely.
For this, we use Lemma \ref{le:vector_convergence}. Under the assumption that $\{a^k\}$ and $\{b^k\}$ are summable, we have that the inequality in (\ref{eq:Theorem_decreasing}) satisfies the relationship in Lemma \ref{le:vector_convergence} with $\bv^k=\left[\sum_{i=1}^{m}\|x_i^k-x_i^*\|^2,\ \sum_{i=1}^m \|v_i^k-\bar v^k\|^2\right]^T$,   $V^k=\left[\begin{array}{cc}
1 &  \kappa_1\gamma^k \cr
0& 1-\kappa_2\gamma^k\cr
\end{array}\right]$, $H^k=\left[\begin{array}{cc}
c^k & 0 \cr
0& 0\cr
\end{array}\right]$, and  $\pi^T=[1, \frac{\kappa_1}{\kappa_2}]^T$. Therefore, according to Lemma \ref{le:vector_convergence}, we know that $\pi^T\bv^k$ converges almost surely, i.e., $\sum_{i=1}^{m}\|x_i^k-x_i^*\|^2+\frac{\kappa_1}{\kappa_2}\sum_{i=1}^m \|v_i^k-\bar v^k\|^2$ converges almost surely. Given that we have proven that $\sum_{i=1}^m \|v_i^k-\bar v^k\|^2$ converges almost surely (see (\ref{eq-sumable})), we have that $\sum_{i=1}^{m}\|x_i^k-x_i^*\|^2$ (or $\|x^k-x^{\ast}\|^2$) converges almost surely. According to Lemma \ref{le:vector_convergence}, we also have $\sum_{ k=0 }^\infty \pi^TH^k\bu^k<\infty$   almost surely, i.e.,
\[
\sum_{ k=0 }^\infty  \left[1, \frac{\kappa_1}{\kappa_2}\right]^T\hspace{-0.15cm}\left[\begin{array}{cc}
c^k & 0 \cr
0& 0\cr
\end{array}\right]\hspace{-0.1cm}\left[\hspace{-0.1cm}\begin{array}{c}
 \left(\phi(x^k)- \phi(x^*)\right)^T(x^k-x^{\ast}) \cr
 0\end{array}\hspace{-0.1cm}\right]\hspace{-0.1cm}<\hspace{-0.1cm}\infty,
\]
or
\begin{equation}\label{eq:phi}
   \sum_{ k=0 }^\infty c^k\left(\phi(x^k)- \phi(x^*)\right)^T(x^k-x^{\ast})<\infty.
\end{equation}

Now using (\ref{eq:phi}) and the proven fact that $\|x^k-x^{\ast}\|^2$ converges  almost surely, we proceed to prove that $x^k$ converges to $x^{\ast}$ almost surely. Because the augmented state decision vector $x^k$ belongs to the compact set $K$ defined in (\ref{eq:K}), we know that the sequence $\{x^k\}$ must have accumulation points in $K$. So the condition   $\sum_{k=0}^{\infty} c^k=\infty$  and (\ref{eq:phi}) mean that there exists a subsequence of $\{x^k\}$, say $\{x^{k_\ell}\}$, along which $\left(\phi(x^k)- \phi(x^*)\right)^T(x^k-x^{\ast})$ converges to zero almost surely.
Recalling that $\phi(\cdot)$ is strictly monotone (see Assumption \ref{as:functions}), one has that the subsequence $\{x^{k_\ell}\}$ must converge to $x^{\ast}$ almost surely.
 This and the fact that $\|x^k-x^{\ast}\|^2$ converges  almost surely imply that
 $x^k$ converges to $x^{\ast}$ almost surely.
\end{proof}

We also need the following Lemma about matrix $L$:
\begin{Lemma 1}\label{Le:rho_2}
Under Assumption \ref{as:L} and a positive sequence $\{\gamma^k\}$ satisfying $
\sum_{k=0}^\infty \gamma^k=\infty$ and $
\sum_{k=0}^\infty (\gamma^k)^2<\infty
$, there always exists a $T>0$ such that when $k\geq T$, we always  have
\[
\|I+\gamma^kL-\frac{{\bf 1}{\bf 1}^T}{m}\|\leq 1-\gamma^k |\rho_2|,
\]
where $\rho_2$ is the   second largest eigenvalue of $L$.
\end{Lemma 1}
\begin{proof}
Under Assumption \ref{as:L}, the matrix $L$ is symmetric,
 so we have that all eigenvalues of $L$ are real numbers.
Since $L$ has non-negative off-diagonal entries and  the diagonal entries $L_{ii}$ are given by
$L_{ii}=-\sum_{j\in\mathbb{N}_i}L_{ij}$, we know that all eigenvalues of $L$ are non-positive (according to the Gershgorin circle theorem), and there is always an eigenvalue equal to 0. Arrange the eigenvalues of $L$ as $\rho_m\leq \rho_{m-1}\leq \cdots\leq \rho_2\leq \rho_1=0$.   It can be verified that the eigenvalues of $I+L$ are given by $1+\rho_m\leq 1+\rho_{m-1}\leq \cdots\leq 1+\rho_2\leq 1+\rho_1=1$, and the eigenvalues of $I+L-\frac{{\bf 1}{\bf 1}^T}{m}$ are given by $\{1+\rho_m,\,1+\rho_{m-1},\, \cdots,\,1+\rho_2,\, 0\}$. Furthermore, the  condition $ \|I+L-\frac{{\bf 1}{\bf 1}^T}{m}\|<1$ in Assumption \ref{as:L} implies that only one eigenvalue of $L$ is zero, and its all  other eigenvalues are strictly less than 0. Hence, we have $\rho_m\leq\rho_{m-1}\leq\cdots\leq \rho_2<0$, i.e., $|\rho_m|\geq|\rho_{m-1}|\geq\cdots\geq |\rho_2|>0$. Since   the eigenvalues of $I+\gamma^kL-\frac{{\bf 1}{\bf 1}^T}{m}$ are  $\{1+\gamma^k\rho_m,\,1+\gamma^k\rho_{m-1},\, \cdots,\,1+\gamma^k\rho_2,\, 0\}$, we have the norm $\|I+\gamma^kL-\frac{{\bf 1}{\bf 1}^T}{m}\|$ being no larger than $|1+\gamma^k\rho_m|$ or $|1+\gamma^k\rho_2|$. Further taking into account the fact that  $\{\gamma^k\}$ is square summable and hence $\gamma^k$ decays to zero, we have that there always exists a $T>0$ such that $|1+\gamma^k \rho_m|=1-\gamma^k|\rho_m|$ and $|1+\gamma^k \rho_2|=1-\gamma^k|\rho_2|$ hold   for $k\geq T$. Given $|\rho_m|\geq  |\rho_2|$,   we have the stated result of the Lemma.
\end{proof}

Using Proposition~\ref{th-main_decreasing}, we are in position to
 establish convergence  of Algorithm 1 assuming that persistent differential-privacy  noise satisfies the following assumption:
\begin{Assumption 1}\label{assumption:dp-noise}
For every $i\in[m]$ and every $k$, conditional on the state $v_i^k$,
the random noise $\zeta_i^k$ satisfies $
\mathbb{E}\left[\zeta_i^k\mid v_i^k\right]=0$ and $\mathbb{E}\left[\|\zeta_i^k\|^2\mid v_i^k\right]=(\sigma_{i}^k)^2$ for all  $k\ge0$, and
\begin{equation}\label{eq:condition_assumption1}
\sum_{k=0}^\infty (\gamma^k)^2\, \max_{i\in[m]}(\sigma_{i}^k)^2 <\infty,
\end{equation} where $\{\gamma^k\}$ is the attenuation sequence from Algorithm 1.
The initial random vectors satisfy
$\mathbb{E}\left[\|v_i^0\|^2\right]<\infty$,  $\forall i\in[m]$.
\end{Assumption 1}

\begin{Remark 1}
Given that $\gamma^k$ decreases with time, (\ref{eq:condition_assumption1}) can be satisfied even when $\{\sigma_i^k\}$ increases with time. For example, under $\gamma^k=\mathcal{O}(\frac{1}{k^{0.9}})$, an increasing $\{\sigma_i^k\}$ with increasing rate no faster than $\mathcal{O}(k^{0.3})$
still satisfies the summable condition in (\ref{eq:condition_assumption1}). Allowing $\{\sigma_i^k\}$ to be increasing with time is key to enabling the strong $\epsilon$-differential privacy in Theorem \ref{th:DP_Algorithm1}.
\end{Remark 1}

\begin{Theorem 1}\label{theorem:convergence_algorithm_1}
Under Assumption \ref{as:functions}, Assumption \ref{as:Lipschitz}, Assumption \ref{as:L},
 and Assumption~\ref{assumption:dp-noise}, if there exists some $T\geq 0$ such that
  $\gamma^k$ and $\lambda^k$ satisfy the following conditions:
 \[
\sum_{k=T}^\infty \gamma^k=\infty, \:\sum_{k=T}^\infty \lambda^k=\infty, \: \sum_{k=T}^\infty (\gamma^k)^2<\infty,\:\sum_{k=T}^\infty \frac{(\lambda^k)^2}{\gamma^k}<\infty,
\]
then Algorithm~1 converges to the Nash equilibrium  of the game in (1)  almost surely.
\end{Theorem 1}

\begin{proof}
The basic idea is to apply
Proposition \ref{th-main_decreasing} to the quantities
$\sum_{i=1}^{m}\| x_i^{k+1}-x_i^*\|^2$ and $\sum_{i=1}^m\|v_i^{k+1}-\bar v^{k+1}\|^2$. Since the results of Proposition \ref{th-main_decreasing} are asymptotic, they
remain valid when the starting index is shifted from $k = 0$ to
$k = T$, for an arbitrary $T\geq0$.
We divide the proof into two parts to analyze $\sum_{i=1}^{m}\|x_i^{k+1}-x_i^*\|^2$ and $\sum_{i=1}^m\|v_i^{k+1}-\bar v^{k+1}\|^2$, respectively.

Part I: We first analyze  $\sum_{i=1}^{m}\|v_i^{k+1}-\bar{v}^{k+1}\|^2$.
For the convenience of analysis, we write the iterates of  $v_i^k$ on per-coordinate expressions. Define for all
$\ell=1,\ldots,d,$ and $k\ge0$,
$v^k(\ell)=\left[[v_1^k]_\ell,\ldots,[v_m^k]_\ell\right]^T$ where $[v_i^k]_\ell$ represents the $\ell$th element of the vector $v_i^k$. Similarly, we  define
$x^k(\ell)=\left[[x_1^k]_\ell,\ldots,[x_m^k]_\ell\right]^T$  and $\zeta^k(\ell) =\left[[\zeta_1^k]_\ell,\ldots,[\zeta_m^k]_\ell\right]^T$. In
this per-coordinate view, (\ref{eq:update_in_Algorithm1})  has the following form  for all $\ell=1,\ldots,d,$
and $k\ge0$:
\begin{equation}\label{eq:update_percord}
\begin{aligned}
v^{k+1}(\ell)&=v^k(\ell)+\gamma^k L v^k(\ell)+\gamma^kL \zeta^k(\ell)  +x^{k+1}(\ell)-x^k(\ell).
\end{aligned}
\end{equation}
Note that the diagonal entries of $L$ are defined as $L_{ii}\triangleq-\sum_{j\in\mathbb{N}_i}L_{ij}$.

The dynamics of the average $v_i^k$, i.e., $\bar{v}^k$, is given by
\begin{equation}\label{eq:bar_v}
\begin{aligned}
&[\bar{v}^{k+1}]_\ell=\frac{{\bf 1}^T}{m} v^{k+1}(\ell)\\
&=\frac{{\bf 1}^T}{m}\left(v^k(\ell)+\gamma^k L v^k(\ell)+\gamma^kL \zeta^k(\ell)  +x^{k+1}(\ell)-x^k(\ell) \right),
\end{aligned}
\end{equation}
where $[\bar{v}^{k+1}]_\ell$ represents the $\ell$-th element of $\bar{v}^{k+1}$.

Under  Assumption \ref{as:L}, we have ${\bf 1}^TL=0$, which simplifies the preceding equation (\ref{eq:bar_v}) to:
\begin{equation}\label{eq:bar_v2}
\begin{aligned}
 [\bar{v}^{k+1}]_\ell= \frac{{\bf 1}^T}{m}\left(v^k(\ell)+   x^{k+1}(\ell)-x^k(\ell) \right),
\end{aligned}
\end{equation}
 where $[\bar{x}^{k+1}]_\ell$ represents the $\ell$-th element of the vector $\bar{x}^{k+1}$.

Combining (\ref{eq:update_percord}), (\ref{eq:bar_v}), and (\ref{eq:bar_v2}) yields
\begin{equation}\label{eq:v_k+1}
\begin{aligned}
&v^{k+1}(\ell)-{\bf 1}[\bar{v}^{k+1}]_{\ell}\\
&=(I+\gamma^kL)v^k(\ell)+\gamma^k L\zeta^k(\ell)+x^{k+1}(\ell)-x^k(\ell)\\
&\quad - \frac{{\bf 1}{\bf 1}^T}{m}\left(v^k(\ell)+   x^{k+1}(\ell)-x^k(\ell) \right)\\
&= \left(I+\gamma^kL-\frac{{\bf 1}{\bf 1}^T}{m}\right)v^k(\ell)+\gamma^k L\zeta^k(\ell)\\
&\quad+\left(I-\frac{{\bf 1}{\bf 1}^T}{m}\right)\left(x^{k+1}(\ell)
- x^k(\ell)\right).
\end{aligned}
\end{equation}

For the sake of notational simplicity, we define
\begin{equation}\label{eq:W^k}
W^k\triangleq I+\gamma^kL-\frac{{\bf 1}{\bf 1}^T}{m},\quad \Pi^k\triangleq I-\frac{{\bf 1}{\bf 1}^T}{m}.
\end{equation}
It can be verified that   $ W^k {\bf 1}=0$ holds, and hence $ W^k {\bf 1}[\bar{v}^k]_\ell=0$ always holds for any $1\leq \ell\leq m$ under Assumption \ref{as:L}. Therefore, (\ref{eq:v_k+1}) can be rewritten as
\begin{equation}\label{eq:v_k+1_3}
\begin{aligned}
v^{k+1}(\ell)-{\bf 1}[\bar{v}^{k+1}]_{\ell}=&W^k\left(v^k(\ell)-{\bf 1}[\bar{v}^k]_\ell\right)+\gamma^k L\zeta^k(\ell)\\
& +\Pi^k \left(x^{k+1}(\ell)
- x^k(\ell)\right),
\end{aligned}
\end{equation}
which further implies
\begin{equation}\label{eq:v_k+1_norm}
\begin{aligned}
&\|v^{k+1}(\ell)-{\bf 1}[\bar{v}^{k+1}]_{\ell} \|^2\\
&=\left\|W^k (v^k(\ell)-{\bf 1}[\bar{v}^k]_\ell )+\Pi^k  (x^{k+1}(\ell)
- x^k(\ell) ) \right\|^2\\
& +2\left\langle W^k (v^k(\ell)-{\bf 1}[\bar{v}^k]_\ell )+\Pi^k  (x^{k+1}(\ell)
- x^k(\ell) ),\, \gamma^k L\zeta^k(\ell) \right\rangle\\
&+\|\gamma^k L\zeta^k(\ell) \|^2\\
&\leq\left\|W^k (v^k(\ell)-{\bf 1}[\bar{v}^k]_\ell )+\Pi^k  (x^{k+1}(\ell)
- x^k(\ell) ) \right\|^2\\
& +2\left\langle W^k (v^k(\ell)-{\bf 1}[\bar{v}^k]_\ell )+\Pi^k  (x^{k+1}(\ell)
- x^k(\ell) ),\, \gamma^k L\zeta^k(\ell) \right\rangle\\
&+(\gamma^k)^2\| L\|^2\zeta^k(\ell) \|^2.
\end{aligned}
\end{equation}

 Taking the conditional expectation, given $\mathcal{F}^k=\{v^0,\,\ldots,v^k\}$, and
using the assumption that the noise $\zeta_i^k$ is with zero mean and variance
$(\sigma_{i}^k)^2$ conditional  on $v_i^k$
(see Assumption~\ref{assumption:dp-noise}),
from the preceding relation we obtain  for all $k\ge0$:
\begin{equation}\label{eq:Ev_k}
\begin{aligned}
&\mathbb{E}\left[\|v^{k+1}(\ell)-{\bf 1}[\bar{v}^{k+1}]_{\ell} \|^2\mathcal{F}^k\right]\\
&\leq\left\|W^k (v^k(\ell)-{\bf 1}[\bar{v}^k]_\ell )+\Pi^k  (x^{k+1}(\ell)
- x^k(\ell) ) \right\|^2\\
&\quad+(\gamma^k)^2\|L\|^2 \mathbb{E}\left[\|\zeta^k(\ell) \|^2\right] \\
&\leq\left(\|W^k \|\| v^k(\ell)-{\bf 1}[\bar{v}^k]_\ell  \|+\|\Pi^k\|\|   x^{k+1}(\ell)
- x^k(\ell)  \|\right)^2\\
&\quad+(\gamma^k)^2\|L\|^2 \mathbb{E}\left[\|\zeta^k(\ell) \|^2\right].
\end{aligned}
\end{equation}
Now we analyze the first term on the right hand side of the preceding inequality. Combined with the  facts that $\|\Pi^k\|=1$ and there exists a $T\geq 0$ such that $0<\|W^k\|\leq 1-\gamma^k |\rho_2|$ holds for $k\geq T$ (see Lemma \ref{Le:rho_2}),  equation (\ref{eq:Ev_k}) implies that there always exists a $T\geq 0$ such that the following inequality always holds for $k\geq T$:
\begin{equation}\label{eq:Ev_k2}
\begin{aligned}
&\mathbb{E}\left[\|v^{k+1}(\ell)-{\bf 1}[\bar{v}^{k+1}]_{\ell} \|^2\mathcal{F}^k\right]\\
&\leq \left((1-\gamma^k|\rho_2|)\|v^k(\ell)-{\bf 1}[\bar{v}^k]_\ell \|+\|x^{k+1}(\ell)
- x^k(\ell) \|\right)^2\\
&\quad+(\gamma^k)^2\|L\|^2\mathbb{E}\left[\|\zeta^k(\ell) \|^2\right].
\end{aligned}
\end{equation}

Using the inequality
$(a+b)^2\le (1+\epsilon) a^2 + (1+\epsilon^{-1})b^2$ valid for any scalars $a,b,$ and $\epsilon>0$, we further have
\begin{equation}\label{eq:Ev_k3}
\begin{aligned}
&\mathbb{E}\left[\|v^{k+1}(\ell)-{\bf 1}[\bar{v}^{k+1}]_{\ell} \|^2\mathcal{F}^k\right]\\
&\leq (1+\epsilon)(1-\gamma^k|\rho_2|)^2\|v^k(\ell)-{\bf 1}[\bar{v}^k]_\ell\|^2\\
&+(1+\epsilon^{-1})\|\  x^{k+1}(\ell)
- x^k(\ell) \|^2 +(\gamma^k)^2\|L\|^2 \mathbb{E}\left[\|\zeta^k(\ell) \|^2\right].
\end{aligned}
\end{equation}
Setting $\epsilon=\frac{\gamma^k|\rho_2|}{1-\gamma^k|\rho_2|}$ (which leads to $(1+\epsilon)=\frac{1}{1-\gamma^k|\rho_2|}$ and $1+\epsilon^{-1}=\frac{1}{\gamma^k|\rho_2|}$) yields
\begin{equation}\label{eq:Ev_k4}
\begin{aligned}
&\mathbb{E}\left[\|v^{k+1}(\ell)-{\bf 1}[\bar{v}^{k+1}]_{\ell} \|^2|\mathcal{F}^k\right]\\
&\leq  (1-\gamma^k|\rho_2|) \|v^k(\ell)-{\bf 1}[\bar{v}^k]_\ell\|^2\\
&+ \frac{1}{\gamma^k|\rho_2|} \| x^{k+1}(\ell)
- x^k(\ell)  \|^2 +(\gamma^k)^2\|L\|^2\mathbb{E}\left[\|\zeta^k(\ell) \|^2\right].
\end{aligned}
\end{equation}
Summing these relations over $\ell=1,\ldots,d$, and noting  $
\sum_{\ell=1}^d \|v^k(\ell) - [\bar v^k]_\ell{\bf
1}\|^2=\sum_{i=1}^m\|v^k_i - \bar v^k \|^2$, $
\sum_{\ell=1}^d \|x^{k+1}(\ell) - x^{k}(\ell)\|^2=\sum_{i=1}^m\|x^{k+1}_i - x^k_i\|^2$, and $\sum_{\ell=1}^d
\|\zeta^k(\ell)\|^2=\sum_{i=1}^m\|\zeta^k_i\|^2$, we obtain
\begin{equation}\label{eq:Ev_k5}
\begin{aligned}
&\mathbb{E}\left[\sum_{i=1}^{m}\|v_i^{k+1}- \bar{v}^{k+1} \|^2\mathcal{F}^k\right]\\
&\leq  (1-\gamma^k|\rho_2|) \sum_{i=1}^{m}\|v_i^k- \bar{v}^k\|^2\\
&+ \frac{1}{\gamma^k|\rho_2|} \sum_{i=1}^{m}\| x_i^{k+1}
- x_i^k \|^2 +(\gamma^k)^2\|L\|^2\mathbb{E}\left[\sum_{i=1}^{m}\|\zeta_i^k  \|^2\right]\\
&\leq  (1-\gamma^k|\rho_2|) \sum_{i=1}^{m}\|v_i^k- \bar{v}^k\|^2\\
&+ \frac{1}{\gamma^k|\rho_2|} \sum_{i=1}^{m}\| x_i^{k+1}
- x_i^k \|^2 +(\gamma^k)^2\|L\|^2 \sum_{i=1}^{m}(\sigma_i^k)^2.
\end{aligned}
\end{equation}
Next, we characterize $\sum_{i=1}^{m}\| x_i^{k+1}
- x_i^k \|^2$. According to (\ref{eq:update_in_Algorithm1}), we have
\begin{equation}\label{eq:x}
\begin{aligned}
\|x_i^{k+1}-x^k_i\|&= \left\|\Pi_{K_i}\left[x_i^k-\lambda^k F_i(x_i^k,v_i^k)\right]-x_i^k\right\|\\
&\leq \left\|  x_i^k-\lambda^k  F_i(x_i^k,v_i^k) -x_i^k\right\|\\
&=\lambda^k\|  F_i(x_i^k,v_i^k)\| \leq \lambda^kC+\lambda^k\tilde{L}\|v_i^k-\bar{v}^k\|,
\end{aligned}
\end{equation}
where in the last inequality we used Lemma \ref{le:bounded}. The preceding inequality further implies
\begin{equation}\label{eq:x2}
\|x_i^{k+1}-x^k_i\|^2\leq 2(\lambda^k)^2C^2+2(\lambda^k)^2\tilde{L}^2\|v_i^k-\bar{v}^k\|^2.
\end{equation}

Plugging (\ref{eq:x2}) into (\ref{eq:Ev_k5}) yields
\begin{equation}\label{eq:Ev_k_final}
\begin{aligned}
&\mathbb{E}\left[\sum_{i=1}^{m}\|v_i^{k+1}- \bar{v}^{k+1} \|^2\mathcal{F}^k\right]\\
&\leq  (1-\gamma^k|\rho_2|) \sum_{i=1}^{m}\|v_i^k- \bar{v}^k\|^2+ \frac{2(\lambda^k)^2\tilde{L}^2}{\gamma^k|\rho_2|} \sum_{i=1}^{m}\| v_i^{k}
- \bar{v}^k \|^2\\
&\quad+ \frac{2m(\lambda^k)^2C^2}{\gamma^k|\rho_2|}  +(\gamma^k)^2\|L\|^2 \sum_{i=1}^{m}(\sigma_i^k)^2.
\end{aligned}
\end{equation}

Part II: Next, we analyze  $\sum_{i=1}^{m}\|x_i^{k+1}-x_i^*\|^2$.

At the Nash equilibrium $ x^{\ast}=[(x_1^{\ast})^T,(x_2^{\ast})^T,\ldots,(x_m^{\ast})^T]^T$, we always have
\[
x_i^{\ast}=\Pi_{K_i}\left[x_i^\ast-\lambda^k F_i(x_i^\ast,\bar{x}^{\ast})\right],
\]
where $\bar{x}^{\ast}=\frac{1}{m}\sum_{i=1}^{m} x_i^{\ast}$.

Therefore, using (\ref{eq:update_in_Algorithm1}), we have
\begin{equation}
\begin{aligned}
&\left\|x_i^{k+1}-x_i^{\ast}\right\|^2\\
&=\left\|\Pi_{K_i}\left[x_i^k-\lambda^k  F_i(x_i^k,v_i^k)\right]-x_i^\ast\right\|^2\\
&= \left\|\Pi_{K_i}\left[x_i^k-\lambda^k  F_i(x_i^k,v_i^k)\right] -\Pi_{K_i}\left[x_i^\ast-\lambda^k  F_i(x_i^\ast,\bar{x}^{\ast})\right]\right\|^2\\
&\leq \left\| x_i^k- x_i^\ast-\lambda^k(  F_i(x_i^k,v_i^k)-   F_i(x_i^\ast,\bar{x}^{\ast})) \right\|^2\\
&\leq \left\| x_i^k- x_i^\ast\right\|^2+(\lambda^k)^2\left\| F_i(x_i^k,v_i^k)-  F_i(x_i^\ast,\bar{x}^{\ast})\right\|^2\\
&\qquad -2\left\langle x_i^k- x_i^\ast,\, \lambda^k( F_i(x_i^k,v_i^k)-  F_i(x_i^\ast,\bar{x}^{\ast}))\right\rangle.
\end{aligned}
\end{equation}

By adding and subtracting   $F_i(x_i^k,\,\bar{v}^k)$ to the inner-product term, we arrive at
\begin{equation}\label{eq:x_ast}
\begin{aligned}
&\left\|x_i^{k+1}-x_i^{\ast}\right\|^2 \\
&\leq \left\| x_i^k- x_i^\ast\right\|^2 + \underbrace{(\lambda^k)^2\left\|  F_i(x_i^k,v_i^k)-  F_i(x_i^\ast,\bar{x}^{\ast})\right\|^2}_{\rm  Term \:1}\\
&\qquad -\underbrace{2\left\langle x_i^k- x_i^\ast,\, \lambda^k( F_i(x_i^k,v_i^k)-   F_i(x_i^k,\bar{v}^k))\right\rangle}_{\rm Term \: 2}\\
&\qquad -\underbrace{2\left\langle x_i^k- x_i^\ast,\, \lambda^k( F_i(x_i^k,\bar{v}^k)-  F_i(x_i^\ast,\bar{x}^{\ast}))\right\rangle}_{\rm Term \: 3}.
\end{aligned}
\end{equation}
Next, we characterize the three terms on the right hand side of (\ref{eq:x_ast}), i.e., Term 1, Term 2, and Term 3, respectively.

Using Lemma \ref{le:bounded}, we can bound Term 1 as follows:
\begin{equation}\label{eq:term1}
\begin{aligned}
{\rm Term \:1}&\leq 2(\lambda^k)^2\left\| F_i(x_i^k,v_i^k)\right\|^2+2(\lambda^k)^2\left\| F_i(x_i^\ast,\bar{x}^{\ast})\right\|^2\\
&\leq 2(\lambda^k)^2\left( C+\tilde{L}\|v_i^k-\bar{v}^k\|\right)^2+2(\lambda^k)^2C^2\\
&\leq 4(\lambda^k)^2 C^2 +4(\lambda^k)^2 \tilde{L}^2\|v_i^k-\bar{v}^k\|^2+2(\lambda^k)^2C^2\\
&=6(\lambda^k)^2 C^2 +4(\lambda^k)^2 \tilde{L}^2\|v_i^k-\bar{v}^k\|^2.
\end{aligned}
\end{equation}

Applying Cauchy–Schwarz inequality to  Term 2 yields
\begin{equation}\label{eq:term2}
\begin{aligned}
&{\rm Term\:2}\\
&\geq -2\lambda^k\|x_i^k- x_i^\ast\| \left\| F_i(x_i^k,v_i^k)-  F_i(x_i^k,\bar{v}^k)\right\|\\
&\geq -\frac{(\lambda^k)^2\|x_i^k- x_i^\ast\|^2}{\gamma^k}
 -\gamma^k\left\| F_i(x_i^k,v_i^k)-  F_i(x_i^k,\bar{v}^k)\right\|^2\\
 &\geq -\frac{(\lambda^k)^2\|x_i^k- x_i^\ast\|^2}{\gamma^k}
 -\gamma^k\tilde{L}^2\left\| v_i^k -\bar{v}^k\right\|^2,
\end{aligned}
\end{equation}
where in the last inequality we used Assumption \ref{as:Lipschitz}, and in the second inequality we used the inequality $2ab\leq \frac{a^2}{\epsilon}+\epsilon b^2$ valid for any $a\in \mathbb{R}$, $b\in\mathbb{R}$, and $\epsilon>0$.

We use Lemma \ref{le:bar_x=bar_v} to treat Term 3. According to Lemma \ref{le:bar_x=bar_v}, we always have $\bar{v}^k=\bar{x}^k$, which further leads to
\begin{equation}\label{eq:term3}
\begin{aligned}
{\rm Term \: 3}&=2\left\langle x_i^k- x_i^\ast,\, \lambda^k(  F_i(x_i^k,\bar{x}^k)-  F_i(x_i^\ast,\bar{x}^{\ast}))\right\rangle\\
&=2\lambda^k \left( F_i(x_i^k,\bar{x}^k)-   F_i(x_i^\ast,\bar{x}^{\ast})\right)^T(x_i^k-x_i^{\ast}).
\end{aligned}
\end{equation}

Plugging (\ref{eq:term1}), (\ref{eq:term2}), and (\ref{eq:term3}) into (\ref{eq:x_ast}) yields
\begin{equation}\label{eq:x_ast_2}
\begin{aligned}
&\left\|x_i^{k+1}-x_i^{\ast}\right\|^2 \\
&\leq \left\| x_i^k- x_i^\ast\right\|^2 + 6(\lambda^k)^2 C^2 +4(\lambda^k)^2 \tilde{L}^2\|v_i^k-\bar{v}^k\|^2\\
&\qquad +\frac{(\lambda^k)^2\|x_i^k- x_i^\ast\|^2}{\gamma^k}
 +\gamma^k\tilde{L}^2\left\| v_i^k -\bar{v}^k\right\|^2\\
&\qquad -2 \lambda^k \left( F_i(x_i^k,\bar{x}^k)-  F_i(x_i^\ast,\bar{x}^{\ast})\right)^T(x_i^k-x_i^{\ast}).
\end{aligned}
\end{equation}

Summing (\ref{eq:x_ast_2}) from $i=1$ to $i=m$ yields
\begin{equation}\label{eq:x_ast_3}
\begin{aligned}
&\sum_{i=1}^{m}\left\|x_i^{k+1}-x_i^{\ast}\right\|^2 \\
&\leq 
\sum_{i=1}^{m}\left\| x_i^k- x_i^\ast\right\|^2 + 6m(\lambda^k)^2 C^2 +4(\lambda^k)^2 \tilde{L}^2\sum_{i=1}^{m}\|v_i^k-\bar{v}^k\|^2\\
&\qquad +\frac{(\lambda^k)^2\sum_{i=1}^{m}\|x_i^k- x_i^\ast\|^2}{\gamma^k}
 +\gamma^k\tilde{L}^2\sum_{i=1}^{m}\left\| v_i^k -\bar{v}^k\right\|^2\\
&\qquad -2 \lambda^k \left(\phi(x^k)- \phi(x^\ast)\right)^T(x^k-x^{\ast}).
\end{aligned}
\end{equation}

From Assumption \ref{as:functions}, we know that $\left(\phi(x^k)- \phi(x^\ast)\right)^T(x^k-x^{\ast})$ in the last term on the right hand side of the proceeding inequality is positive for all $x^k\neq x^{\ast}$.

Next, we combine Step I and Step II to prove the theorem.

Defining $\bv^k=\left[\sum_{i=1}^{m}\|x_i^k-x_i^*\|^2,\ \sum_{i=1}^m \|v_i^k-\bar v^k\|^2\right]^T$, we have the following relations from  (\ref{eq:Ev_k_final}) and (\ref{eq:x_ast_3}):
\begin{equation}\label{eq:stacked}
\begin{aligned}
\mathbb{E}\left[\bv^{k+1}|\mathcal{F}^k\right]\leq(V^k+A^k)\bv^k-2\lambda^k\Phi^k+B^k,
\end{aligned}
\end{equation}
where

\[
\begin{aligned}
&V^k=\left[\begin{array}{cc}
1 & {\tilde{L}^2}\gamma^k \cr
0& 1-\gamma^k|\rho_2|
\end{array}\right],\\
&A^k=\left[\begin{array}{cc}
\frac{(\lambda^k)^2}{\gamma^k} &
4(\lambda^k)^2\tilde{L}^2\cr
 0& \frac{2(\lambda^k)^2\tilde{L}^2}{\gamma^k|\rho_2|} \end{array}\right],\\
&\Phi^k= \left[\begin{array}{c}
 \left(\phi(x^k)- \phi(x^*)\right)^T(x^k-x^{\ast}) \cr
 0\end{array}\right] ,\\
 & B^k=\left[\begin{array}{c}  6m(\lambda^k)^2 C^2
\cr
\frac{2m(\lambda^k)^2C^2}{\gamma^k|\rho_2|}  +(\gamma^k)^2\|L\|^2 \sum_{i=1}^{m}(\sigma_i^k)^2\end{array}\right].
\end{aligned}
\]

Using Assumption~\ref{assumption:dp-noise} and the conditions of the theorem  $\sum_{k=T}^\infty (\gamma^k)^2<\infty$ and $\sum_{k=T}^\infty \frac{(\lambda^k)^2}{\gamma^k}<\infty$, we have that all elements of the matrices of $A^k$ and $B^k$ are summable. Therefore,  we  have $ \sum_{i=1}^{m}\|x_i^k-x_i^*\|^2 $ and $ \sum_{i=1}^m \|v_i^k-\bar v^k\|^2 $ satisfying the conditions of Proposition~\ref{th-main_decreasing} with
$\kappa_1= \tilde{L}^2 $, $\kappa_2=|\rho_2|$, $c^k=2\lambda^k$,
$a^k=\max\{\frac{(\lambda^k)^2}{\gamma^k}, \,
4(\lambda^k)^2\tilde{L}^2,\, \frac{2(\lambda^k)^2\tilde{L}^2}{\gamma^k|\rho_2|} \}$, and $b^k= \max\{ 6m(\lambda^k)^2 C^2,\,
\frac{2m(\lambda^k)^2C^2}{\gamma^k|\rho_2|}  +(\gamma^k)^2\|L\|^2 \sum_{i=1}^{m}(\sigma_i^k)^2\}$.
\end{proof}
\begin{Remark 1}
  The conditions for $\{\gamma^k\}$ and $\{\lambda^k\}$ can be satisfied, e.g., by setting $\lambda^k=\frac{c_1}{1+c_2k}$ and $\gamma^k=\frac{c_3}{1+c_4k^\varrho}$ with any $0.5<\varrho<1$,  $c_1>0,\,c_2>0,\,c_3>0$, and $c_4>0$.
\end{Remark 1}

\begin{Remark 1}
   In the derivation, it can be seen that the   aggregate-estimation error $\sum_{i=1}^{m}\|v_i^k-\bar{v}^k\|$ does not decrease geometrically  with $k$, which makes it impossible to use existing proof techniques for distributed Nash-equilibrium computation algorithms. In fact, in existing distributed Nash-equilibrium computation algorithms (e.g., \cite{koshal2016distributed,belgioioso2020distributed,parise2019distributed,gadjov2020single,zhu2022asynchronous,liang2017distributed}) and their stochastic variants (e.g., \cite{lei2022distributed} and \cite{wang2022differentially}), because the inter-player interaction is persistent, the  aggregative-estimation  error $\sum_{i=1}^{m}\|v_i^k-\bar{v}^k\|$ always decreases geometrically, which makes it possible to separate the evolution analysis of the aggregate-estimation error and the   decision distance from the Nash equilibrium. However, in the proposed algorithm, the diminishing $\gamma^k$ leads to a non-geometric decreasing of the  aggregative estimation error, which makes it impossible to analyze the evolution of the aggregate estimate $v_i^k$ and the decision $x_i^k$ separately, and hence makes the proposed proof technique fundamentally different from existing analysis.
\end{Remark 1}

\begin{Remark 1}
Communication imperfections  can be modeled as channel noises, which can be regarded as the differential-privacy noise   here.  Therefore, Algorithm 1 can also  counteract such communication imperfections in distributed equilibrium computation of aggregative games.
\end{Remark 1}

\begin{Remark 1}\label{Re-rate}
Because the evolution of $x_i^k$ to the Nash equilibrium  satisfies the conditions in Proposition \ref{th-main_decreasing}, we can leverage Proposition \ref{th-main_decreasing} to examine the convergence speed.
The relation~\eqref{eq:Theorem_decreasing} implies that the following inequality holds almost surely for all $k\ge0$:
\begin{equation}
\begin{aligned}
&\mathbb{E}\left[\sum_{i=1}^{m}\|x_i^{k+1}-x_i^*\|^2|\mathcal{F}^k\right]
\le  (1+a^k)\sum_{i=1}^{m}\|  x_i^k -x_i^*\|^2 \cr
&+\left( \kappa_1\gamma^k +a^k\right) \sum_{i=1}^m \|v_i^k - \bar v^k\|^2 + b^k.
\end{aligned}
\end{equation}
 The first relationship in (\ref{eq-sumable}) (i.e., $\sum_{k=0}^\infty \kappa_2\gamma^k\sum_{i=1}^m \|v_i^k - \bar v^k\|^2<\infty$)  implies that  $\left( \kappa_1\gamma^k +a^k\right)\sum_{i=1}^m \|v_i^k - \bar v^k\|^2$ is summalbe, where we used the fact that  $\{a^k\}$ is summable and hence decreases faster than $\{\gamma^k\}$ (which is not summable). Therefore, we have the following relationship for the error $\sum_{i=1}^{m}\|x_i^{k+1}-x_i^*\|^2$ almost surely:
 \begin{equation}
\begin{aligned}
&\mathbb{E}\left[\sum_{i=1}^{m}\|x_i^{k+1}-x_i^*\|^2|\mathcal{F}^k\right]
\le  (1+a^k)\sum_{i=1}^{m}\| x_i^k -x_i^*\|^2  +  \hat{b}^k,
\end{aligned}
\end{equation}
where the sequence of $\hat{b}^k=\left( \kappa_1\gamma^k +a^k\right) \sum_{i=1}^m \|v_i^k - \bar v^k\|^2 + b^k$ is summable.
Further using the fact that $a^k$ is also a summable non-negative sequence, and all non-negative summable sequences decrease to zero with a rate no slower than $\mathcal{O}(\frac{1}{k})$, we have that  $\mathbb{E}\left[\sum_{i=1}^{m}\|x_i^{k+1}-x_i^*\|^2\right]$ converges with a rate no slower than $\mathcal{O}(\frac{1}{k})$.
Moreover, from (\ref{eq:Ev_k_final}), it can be seen that the decreasing speed of  $\sum_{i=1}^{m}\|v_i^k-  \bar{v}^{k}\|^2 $ increases with an increase in $|\rho_2|$, which corresponds to the  second largest eigenvalue of $L$. Therefore, the decreasing speed of  $\sum_{i=1}^{m}\|v_i^{k}-   \bar{v}^{k}\|^2 $ to zero increases with an increase in the   absolute value of the second largest eigenvalue of $L$ in Assumption \ref{as:L}.
\end{Remark 1}

\subsection{Privacy Analysis for Algorithm 1}\label{se:privacy_Algorithm1}
Following the idea of differential-privacy design for distributed optimization in \cite{Huang15}, 
we  define the sensitivity of a distributed  Nash-equilibrium computation algorithm to problem (\ref{eq:formulation}) as follows:
\begin{Definition 1}\label{de:sensitivity}
  At each iteration $k$, for any initial state $\vartheta^0$ and any adjacent distributed games  $\mathcal{P}$ and $\mathcal{P'}$,  the sensitivity of a Nash-equilibrium computation algorithm is
  \begin{equation}
  \Delta^k\triangleq \sup\limits_{\mathcal{O}\in\mathbb{O}}\left\{\sup\limits_{\vartheta\in\mathcal{R}^{-1}(\mathcal{P},\mathcal{O},\vartheta^0),\:\vartheta'\in\mathcal{R}^{-1}(\mathcal{P'},\mathcal{O},\vartheta^0)}\hspace{-0.3cm}\|\vartheta^{k+1}-\vartheta'^{k+1}\|_1\right\}.
  \end{equation}
\end{Definition 1}

Then, we have the following lemma:
\begin{Lemma 1}\label{Le:Laplacian}
In Algorithm 1, at each iteration $k$, if each player adds a noise vector $\zeta_i^k\in\mathbb{R}^d$ consisting of $d$ independent Laplace noises with  parameter $\nu^k$ to shared messages $v_i^k$ such that $\sum_{k=1}^{T_0}\frac{\Delta^k}{\nu^k}\leq \bar\epsilon$, then the iterative distributed Algorithm 1 is $\epsilon$-differentially private with the cumulative privacy budget for iterations from $k=0$ to $k=T_0$ less than $\bar\epsilon$.
\end{Lemma 1}
\begin{proof}
The lemma can be obtained following the same line of reasoning of Lemma 2 in  \cite{Huang15} (also see Theorem 3 in \cite{ye2021differentially}).
\end{proof}

As indicated in \cite{Huang15}, since the change in the payoff function $f_i$ can be arbitrarily large in the adjacency definition in Definition \ref{de:adjacency}, we have to make the following assumption to ensure bounded sensitivity:
\begin{Assumption 1}\label{as:bounded_gradients}
 The pseudo-gradients $F_i(x_i^k,v_i^k)$ of all individual players are bounded, i.e., there exists a constant $\bar{C}$ such that $\|F_i(x_i^k,v_i^k)\|_1\leq \bar{C}$ holds for all $x_i^k,v_i^k\in\mathbb{R}^d$ and $1\leq i \leq m$.
\end{Assumption 1}
\begin{Remark 1}
  Note that this uniformly bounded condition on $F_i(x_i^k,v_i^k)$ is only required for privacy analysis. As indicated in Remark \ref{re:boundedness}, it is not needed in our convergence analysis.
\end{Remark 1}
\begin{Theorem 1}\label{th:DP_Algorithm1}
Under Assumptions \ref{as:functions}, \ref{as:Lipschitz} \ref{as:L},  \ref{as:bounded_gradients}, if  $\{\lambda^k\}$ and $\{\gamma^k\}$ satisfy the conditions in Theorem \ref{theorem:convergence_algorithm_1}, and all elements of $\zeta_i^k$ are drawn independently from  Laplace distribution ${\rm Lap}(\nu^k)$ with $(\sigma_i^k)^2=2(\nu^k)^2$ satisfying Assumption \ref{assumption:dp-noise}, then all players will converge almost surely to the Nash equilibrium. Moreover, 
\begin{enumerate}
\item Algorithm 1 is  $\epsilon$-differentially private with the cumulative privacy budget bounded by $\epsilon\leq \sum_{k=1}^{T_0}\frac{2\bar{C}\lambda^k}{\nu^k}$ for iterations from $k=0$ to $k=T_0$ where $\bar{C}$ is from Assumption \ref{as:bounded_gradients}. And the cumulative privacy budget is always finite for $T_0\rightarrow\infty$  when the sequence  $\{\frac{\lambda^k}{\nu^k}\}$ is summable;
\item Suppose that two non-negative sequences  $\{\nu'^k\}$ and $\{\lambda^k\}$ have a finite sequence-ratio sum $\Phi_{\lambda,\nu'}\triangleq \sum_{k=1}^{\infty}\frac{\lambda^k}{\nu'^k}$. Then setting the Laplace noise parameter $\nu^k$ as $\nu^k=\frac{2 \bar{C}\Phi_{\lambda,\nu'}}{ \epsilon  }\nu'^k$ ensures that Algorithm 1 is $\epsilon$-differentially private with any cumulative privacy budget $\epsilon>0$ even when the number of iterations goes to infinity;
\item In the special case where $\lambda^k=\frac{1}{k}$ and $\gamma^k=\frac{1}{k^{0.9}}$,   setting $\nu^k=\frac{2\bar{C} \Phi}{\epsilon}k^{0.3}$ with  $\Phi\triangleq \sum_{k=1}^{\infty}\frac{1}{k^{1.3}}\approx 3.93$ (which can be verified to satisfy Assumption \ref{assumption:dp-noise}) ensures that  Algorithm 1 is always $\epsilon$-differentially private for any cumulative privacy budget $ \epsilon>0$ even when the number of iterations goes to infinity.
\end{enumerate}
\end{Theorem 1}
\begin{proof}
Because the Laplace noise satisfies Assumption \ref{assumption:dp-noise}, it follows from  Theorem \ref{theorem:convergence_algorithm_1} that the iterate $x_i^k$ of every player $i$  will converge to the Nash equilibrum $x_i^{\ast}$ almost surely.

To prove the three statements on the strength of differential privacy, we first prove that the sensitivity of the algorithm satisfies $\Delta^k\leq 2\bar{C}\lambda^k$. Given two adjacent distributed games $\mathcal{P}$ and $\mathcal{P'}$, for any given fixed observation $\mathcal{O}$ and initial state $\vartheta^0=\left[(x^0)^T,\,(v^0)^T\right]^T$, the sensitivity is determined by $\|\mathcal{R}^{-1}(\mathcal{P},\mathcal{O},\vartheta^0)-\mathcal{R}^{-1}(\mathcal{P'},\mathcal{O},\vartheta^0)\|_1$ according to Definition \ref{de:sensitivity}.
Since in $\mathcal{P}$ and $\mathcal{P'}$, there is only one payoff function that is different, we  represent this different payoff function as the  $i$th one, i.e., $f_i(\cdot)$, without loss of generality. Because the observations under $\mathcal{P}$ and $\mathcal{P'}$  are identical, we have
\[
 v_j^k={v'}_j^k,\quad \forall k\geq 0,\quad \forall j\neq i.
\]
The preceding relationship implies
\[
x_j^{k+1}-x_j^k={x'}_j^{k+1}-{x'}_j^{k},\quad \forall k\geq 0,\quad  \forall j\neq i
\]
according to the update rule in (\ref{eq:update_in_Algorithm1}), and further
\[
x_j^{k}={x'}_j^k,\quad \forall j\neq i
\]
because of the identical initial condition $x^0={x'}^0$.

 Therefore,  we have the following relationship for the sensitivity  of  Algorithm 1:
\[
\begin{aligned}
&\|\mathcal{R}^{-1}(\mathcal{P},\mathcal{O},\vartheta^0)-\mathcal{R}^{-1}(\mathcal{P'},\mathcal{O},\vartheta^0)\|_1\\
&=\left\| \left[\begin{array}{c}x^{k+1}\\v^{k+1}\end{array}\right] -\left[\begin{array}{c}{x'}^{k+1}\\{v'}^{k+1}\end{array}\right]\right\|_1
=\left\| \left[\begin{array}{c}x^{k+1}-{x'}^{k+1}\\v^{k+1}-{v'}^{k+1}\end{array}\right] \right\|_1\\
&=\left\| \left[\begin{array}{c}x^{k+1}-{x'}^{k+1}\\0\end{array}\right] \right\|_1=\left\| x^{k+1}-{x'}^{k+1} \right\|_1,
\end{aligned}
\]
where we used the fact that the observations $v^{k+1}$ and ${v'}^{k+1}$ are the same in the second last equality.

Using the update rule in (\ref{eq:update_in_Algorithm1}), and the fact that there is just one player that has different payoff functions in adjacent games (represented as the $i$th one), we can further write the above relationship as
\[
\begin{aligned}
&\|\mathcal{R}^{-1}(\mathcal{P},\mathcal{O},\vartheta^0)-\mathcal{R}^{-1}(\mathcal{P'},\mathcal{O},\vartheta^0)\|_1\\
&=\left\| \Pi_{K_i}\left[x_i^{k}-\lambda^{k} F_i(x_i^{k},v_i^{k})\right]\hspace{-0.1cm} -\hspace{-0.1cm}\Pi_{K_i}\left[{x'}_i^{k}-\lambda^{k} {F'}_i({x'}_i^{k},{v'}_i^{k})\right] \right\|_1 \\
&\leq\left\|  x_i^{k}-\lambda^{k} F_i(x_i^{k},v_i^{k})  -  {x'}_i^{k}-\lambda^{k} {F'}_i({x'}_i^{k},{v'}_i^{k}) \right\|_1 \\
& =\left\|  \lambda^{k} F_i(x_i^{k},v_i^{k})  -   \lambda^{k} {F'}_i({x'}_i^{k},{v'}_i^{k}) \right\|_1,
\end{aligned}
\]
where we have used the relationship $x^{k}={x'}^k$ that holds for all  $k\geq 0$ in the last equality.

According to Assumption \ref{as:bounded_gradients}, we have
\[
\|F_i(x_i^{k},v_i^{k})\|_1\leq \bar{C},\quad \|{F'}_i({x'}_i^{k},{v'}_i^{k})\|_1\leq \bar{C}.
\]
Combining the two preceding relations leads to
\[
\begin{aligned}
&\|\mathcal{R}^{-1}(\mathcal{P},\mathcal{O},\vartheta^0)-\mathcal{R}^{-1}(\mathcal{P'},\mathcal{O},\vartheta^0)\|_1 \leq 2\lambda^{k}\bar{C}.
\end{aligned}
\]

Using Lemma \ref{Le:Laplacian}, we can  obtain that the cumulative privacy budget is always less than $ \sum_{k=1}^{T_0}\frac{2\bar{C}\lambda^k}{\nu^k}$. Hence, the cumulative privacy budget $\epsilon$ will always be finite even when the number of iterations $T_0$ tends to infinity if  the sequence $\{\frac{\lambda^k}{\nu^k}\}$ is summable, i.e.,  $\sum_{k=0}^{\infty}\frac{\lambda^k}{\nu^k}<\infty$.

According to the obtained result that the cumulative privacy budget is inversely proportional to $\nu^k$, we can obtain the second statement from the first statement by  scaling $\nu^k$ proportionally. The result in the third statement can be obtained by specializing the selection of $\lambda^k$, $\gamma^k$, and $\nu^k$ sequences.
\end{proof}

Note that to ensure that the cumulative differential-privacy budget is finite (an unbounded privacy budget means complete loss of privacy protection),  \cite{ye2021differentially} and \cite{Huang15} have to use a summable stepsize  (geometrically-decreasing stepsize, more specifically), which, however, also makes it impossible to ensure  convergence to the exact desired equilibrium. In our approach, by allowing the stepsize sequence to be non-summable, we achieve both accurate convergence and finite cumulative privacy budget, even when the number of iterations goes to infinity.   In fact, to our knowledge, this is the first time that almost-sure convergence to a Nash equilibrium is achieved under rigorous $\epsilon$-differential privacy even with the number of iterations going to infinity.

\begin{Remark 1}
  It is worth noting that  to ensure the boundedness of the cumulative privacy budget $\epsilon=\sum_{k=1}^{\infty}\frac{2\bar{C}\lambda^k}{\nu^k}$  when  $k\rightarrow \infty$, our algorithm uses   Laplace noise with parameter $\nu^k$  increasing with time (since we require  $\frac{\lambda^k}{\nu^k}$ to be summable while  $\{\lambda^k\}$ is non-summable).  Because the strength of shared signal is always  $v_i^k$, an increasing $\nu^k$  makes the relative level between noise $\zeta_i^k$ and signal $v_i^k$ increase  with time. However, since what actually  feeds into the algorithm is $\gamma^k{\rm Lap}(\nu^k)$, and  the increase  in the noise level  $\nu^k$ is outweighed by the decrease of $\gamma^k$ (see  Assumption \ref{assumption:dp-noise}), the actual noise fed into the algorithm  still decays with time, which makes it possible for  Algorithm 1 to ensure every player's  almost sure convergence to the Nash equilibrium. Moreover, according to Theorem \ref{theorem:convergence_algorithm_1}, such almost sure convergence is not affected by scaling  $\nu^k$ by any constant coefficient $\frac{1}{\epsilon}>0$ so as to achieve any desired level of $\epsilon$-differential privacy, as long as the Laplace noise parameter $\nu^k$ (with associated variance $(\sigma_i^k)^2=2(\nu^k)^2$) satisfies Assumption \ref{assumption:dp-noise}. 
\end{Remark 1}

\section{Extension to stochastic aggregative games}\label{se:algorithm2}
In this section,  we prove that the proposed distributed algorithm can ensure the almost sure convergence of all agents to the Nash equilibrium even when individual agents only have access to a stochastic estimate of their payoff functions. Such  stochastic Nash-equilibrium computing problems arise frequently in practical applications like electricity markets \cite{franci2021stochastic,yousefian2015self} and transportation systems \cite{watling2006user} where the payoff functions are subject to stochastic uncertainties.

Representing the stochastic version of the payoff functions as $f_i(x_i,\bar{x},\xi_i)$ for player $i$, where $\bar{x}\triangleq \frac{\sum_{i=1}^{m}x_i}{m}$, and $\xi_i\in\mathbb{R}^d$ is a random vector, we can formulate the stochastic game that player $i$ faces as the following parameterized optimization problem:
\begin{equation}\label{eq:formulation_stochastic}
 \min \mathbb{E}\left[f_i(x_i,\bar{x},\xi_i)\right]\quad {\rm s.t.}\quad x_i\in K_i\:\: {\rm and}\:\: \bar{x}\in\bar{K},
\end{equation}
where the expected value is taken with respect to $\xi_i$.
The constraint set $K_i$ and the function $f_i(\cdot)$ are assumed to be known to player $i$ only.

When the payoff functions  are given through the expectation,
the pseudo-gradients that individual players can access become stochastic, i.e.,
the gradient mapping $F(x,\bar{x})$ has components
\[F_i(x,\bar{x})=\mathbb{E}\left[\nabla_{x_i} f_i(x_i,\bar{x},\xi_i)\right], \quad \forall  i\in[m].\]
In this case, in Algorithm 1, the mapping $F_i(x_i^k,v_i^k)$ is replaced with a sampled mapping
\[\tilde F_i(x_i^k,v_i^k,\xi^k_i)= \nabla_{x_i} f_i(x_i^k,v_i^k,\xi^k_i),\quad \forall  i\in[m].\] Accordingly, our privacy-preserving distributed algorithm reduces to:

\noindent\rule{0.49\textwidth}{0.5pt}
\noindent\textbf{Algorithm 2: Differentially-private distributed algorithm for stochastic aggregative games with guaranteed convergence}

\noindent\rule{0.49\textwidth}{0.5pt}
\begin{enumerate}[wide, labelwidth=!, labelindent=0pt]
    \item[] Parameters: Stepsize $\lambda^k>0$ and
    weakening factor $\gamma^k>0$.
    \item[] Every player $i$ maintains one decision variable $x_i^k$, which is initialized with a random vector in $K_i\subseteq\mathbb{R}^d$, and an estimate of the aggregate decision $v_i^k$, which is initialized as $v_i^0=x_i^0$.
    \item[] {\bf for  $k=1,2,\ldots$ do}
    \begin{enumerate}
        \item Every player $j$ adds persistent differential-privacy noise   $\zeta_j^{k}$ 
        to its estimate
    $v_j^k$,  and then sends the obscured estimate $v_j^k+\zeta_j^{k}$ to agent
        $i\in\mathbb{N}_j$.
        \item After receiving  $v_j^k+\zeta_j^k$ from all $j\in\mathbb{N}_i$, player $i$ updates its decision variable and estimate  as follows:
        \begin{equation}\label{eq:update_in_Algorithm2}
        \begin{aligned}
             x_i^{k+1}&=\Pi_{K_i}\left[x_i^k-\lambda^k\nabla \tilde{F}_i(x_i^k,v_i^k,\xi_i^k)\right],\\
             v_i^{k+1}&=v_i^k+\gamma^k\sum_{j\in \mathbb{N}_i} L_{ij}(v_j^k+\zeta_j^k-v_i^k-\zeta_i^k)+x_i^{k+1}-x_i^k.
        \end{aligned}
        \end{equation}
                \item {\bf end}
    \end{enumerate}
\end{enumerate}
\vspace{-0.1cm} \rule{0.49\textwidth}{0.5pt}

\subsection{Convergence Analysis}
Next, we prove that Algorithm 2 can ensure the convergence of the decision vector $x^k\triangleq[(x_1^k)^T,\cdots,(x_m^k)^T]^T$  to the exact Nash equilibrium point $x^{\ast}\triangleq[(x_1^\ast)^T,\cdots,(x_m^\ast)^T]^T$, even in the presence of differential-privacy noise $\zeta_i^k$ and stochastic pseudo-gradient $\tilde{F}_i(x_i,v_i^k,\xi_i^k)$.
To this end, similar to \cite{yousefian2015self}, we first formalize the noise in pseudo-gradients:

 \begin{Assumption 1}\label{As:unbiased}
 Let $\mathcal{F}^k\triangleq\{\xi^0,\cdots,\xi^k\}$ be the family of sigma algebra with $\xi^k=[(\xi_1^k)^T,\,\cdots,\,(\xi_m^k)^T]^T$, we have the following relationship almost surely:
\begin{equation}\label{eq:unbiased}
\mathbb{E}\left[\tilde{F}_i(x_i^k,v_i^k,\xi_i^k)|\mathcal{F}^k\right]=F_i(x_i^k,v_i^k),
\end{equation}
\begin{equation}\label{eq:variance}
\mathbb{E}\left[\left\|\tilde{F}_i(x_i^k,v_i^k,\xi_i^k)-F_i(x_i^k,v_i^k)\right\|^2|\mathcal{F}^k\right]\leq (\mu^k)^2,
\end{equation}
where $\mu^k$ is some positive scalar.
 \end{Assumption 1}

\begin{Theorem 2}\label{theorem:convergence_algorithm_2}
Under Assumption \ref{as:functions}, Assumption \ref{as:Lipschitz}, Assumption \ref{as:L},
  Assumption~\ref{assumption:dp-noise}, and Assumption \ref{As:unbiased}, if there exists some $T\geq 0$ such that for all $k\geq T$,
  $\gamma^k$ and $\lambda^k$ satisfy the following conditions:
\[
\sum_{k=T}^\infty \gamma^k=\infty, \:\sum_{k=T}^\infty \lambda^k=\infty, \: \sum_{k=T}^\infty (\gamma^k)^2<\infty,\:\sum_{k=T}^\infty \frac{(\lambda^k)^2}{\gamma^k}<\infty,\] and $\sum_{k=T}^\infty (\lambda^k\mu^k)^2<\infty$,
then Algorithm~2 converges to the Nash equilibrium   of the game in~(\ref{eq:formulation_stochastic})  almost surely.
\end{Theorem 2}
\begin{proof}
Similar to the proof of Theorem \ref{theorem:convergence_algorithm_1}, the basic idea is still to apply
Proposition \ref{th-main_decreasing} to the quantities
$\sum_{i=1}^{m}\| x_i^{k+1}-x_i^*\|^2$ and $\sum_{i=1}^m\|v_i^{k+1}-\bar v^{k+1}\|^2$. Since the stochasticity in $\tilde{F}_i(x_i,\bar{x},\xi_i^k)$ does not affect the dynamics of $v_i^k$, the relationship for  $\sum_{i=1}^m\|v_i^{k+1}-\bar v^{k+1}\|^2$ in Algorithm 1 still holds under Algorithm 2, i.e., we still have
\begin{equation}\label{eq:Ev_k_final_algorithm2}
\begin{aligned}
&\mathbb{E}\left[\sum_{i=1}^{m}\|v_i^{k+1}- \bar{v}^{k+1} \|^2\mathcal{F}^k\right]\\
&\leq  (1-\gamma^k|\rho_2|) \sum_{i=1}^{m}\|v_i^k- \bar{v}^k\|^2+ \frac{2(\lambda^k)^2\tilde{L}^2}{\gamma^k|\rho_2|} \sum_{i=1}^{m}\| v_i^{k}- \bar{v}^k \|^2\\
&
\quad+ \frac{2m(\lambda^k)^2C^2}{\gamma^k|\rho_2|}  +(\gamma^k)^2\|L\|^2 \sum_{i=1}^{m}(\sigma_i^k)^2
\end{aligned}
\end{equation}
for $\mathcal{F}^k=\{x^0,\,v^0,\cdots,x^k,\,v^k\}$.

Therefore, we only characterize $\sum_{i=1}^{m}\|x_i^{k+1}-x_i^*\|^2$, whose evolution is affected by the replacement of $F_i(x_i^k,v_i^k)$ with $\tilde{F}_i(x_i^k,v_i^k,\xi_i^k)$.

Using the relation $x_i^{\ast}=\Pi_{K_i}\left[x_i^\ast-\lambda^k F_i(x_i^\ast,\bar{x}^{\ast})\right]
$ with $\bar{x}^{\ast}=\frac{1}{m}\sum_{i=1}^{m}x_i^{\ast}$, from (\ref{eq:update_in_Algorithm2}), we can arrive at
\begin{equation}\label{eq:x_ast_algorithm2}
\begin{aligned}
&\left\|x_i^{k+1}-x_i^{\ast}\right\|^2\\
&=\left\|\Pi_{K_i}\left[x_i^k-\lambda^k \tilde{F}_i(x_i^k,v_i^k,\xi_i^k)\right]-x_i^\ast\right\|^2\\
&= \left\|\Pi_{K_i}\hspace{-0.05cm}\left[x_i^k-\lambda^k  \tilde{F}_i(x_i^k,v_i^k,\xi_i^k)\right]\hspace{-0.1cm}-\hspace{-0.1cm}\Pi_{K_i}\hspace{-0.05cm}\left[x_i^\ast-\lambda^k  F_i(x_i^\ast,\bar{x}^{\ast})\right]\right\|^2\\
&\leq \left\| x_i^k- x_i^\ast-\lambda^k(  \tilde{F}_i(x_i^k,v_i^k,\xi_i^k)-   F_i(x_i^\ast,\bar{x}^{\ast})) \right\|^2\\
&\leq \left\| x_i^k- x_i^\ast\right\|^2+(\lambda^k)^2\left\|  \tilde{F}_i(x_i^k,v_i^k,\xi_i^k)-   F_i(x_i^\ast,\bar{x}^{\ast})\right\|^2\\
&\qquad -2\left\langle x_i^k- x_i^\ast,\, \lambda^k(  \tilde{F}_i(x_i^k,v_i^k,\xi_i^k)-  F_i(x_i^\ast,\bar{x}^{\ast}))\right\rangle.
\end{aligned}
\end{equation}

For the second term on the right hand side of the above inequality,  we can bound it by adding and subtracting $F_i(x_i^k,v_i^k)$:
\begin{equation}\label{eq:bound_1}
\begin{aligned}
&\left\| \tilde{F}_i(x_i^k,v_i^k,\xi_i^k)-   F_i(x_i^\ast,\bar{x}^{\ast})\right\|^2\\
=&\left\|  \tilde{F}_i(x_i^k,v_i^k,\xi_i^k)-  F_i(x_i^k,v_i^k) +  F_i(x_i^k,v_i^k) -  F_i(x_i^\ast,\bar{x}^{\ast})\right\|^2\\
\leq& 2\left\|  \tilde{F}_i(x_i^k,v_i^k,\xi_i^k)-   F_i(x_i^k,v_i^k)\right\|^2   \\
&+2\left\|  F_i(x_i^k,v_i^k)-  F_i(x_i^\ast,\bar{x}^{\ast})\right\|^2.
\end{aligned}
\end{equation}

Plugging (\ref{eq:bound_1}) into (\ref{eq:x_ast_algorithm2}) yields
\begin{equation}\label{eq:x_ast_algorithm2_2}
\begin{aligned}
&\left\|x_i^{k+1}-x_i^{\ast}\right\|^2\\
&\leq \left\| x_i^k- x_i^\ast\right\|^2+2(\lambda^k)^2\left\|  \tilde{F}_i(x_i^k,v_i^k,\xi_i^k)-  F_i(x_i^k,v_i^k)\right\|^2\\
&\quad + 2(\lambda^k)^2\left\|  F_i(x_i^k,v_i^k)-  F_i(x_i^\ast,\bar{x}^{\ast})\right\|^2\\
&\quad -2\left\langle x_i^k- x_i^\ast,\, \lambda^k( \tilde{F}_i(x_i^k,v_i^k,\xi_i^k)-  F_i(x_i^\ast,\bar{x}^{\ast}))\right\rangle.
\end{aligned}
\end{equation}

Taking the conditional expectation, given $\mathcal{F}^k=\{v^0,\, x^0,\,\ldots,v^k,\,x^k\,\}$,
from the preceding relation we obtain  for all $k\ge0$:
\begin{equation}\label{eq:x_ast_algm2}
\begin{aligned}
&\mathbb{E}\left[\left\|x_i^{k+1}-x_i^{\ast}\right\|^2 | \mathcal{F}^k\right]\leq \left\| x_i^k- x_i^\ast\right\|^2 +2(\lambda^k\mu^k)^2\\
&\quad + 2(\lambda^k)^2\left\|  F_i(x_i^k,v_i^k)-  F_i(x_i^\ast,\bar{x}^{\ast})\right\|^2\\
&\quad -2\left\langle x_i^k- x_i^\ast,\, \lambda^k(  F_i(x_i^k,v_i^k)-   F_i(x_i^\ast,\bar{x}^{\ast}))\right\rangle,
\end{aligned}
\end{equation}
where we used  the assumption that   $ \tilde{F}_i(x_i^k,v_i^k,\xi_i^k)$ is an unbiased estimate of $ F_i(x_i^k,v_i^k)$ with  variance $(\mu^k)^2$ (see Assumption~\ref{As:unbiased}).

By adding and subtracting   $F_i(x_i^k,\,\bar{v}^k)$ to the inner-product term, we arrive at
\begin{equation}\label{eq:x_ast_algo2_2}
\begin{aligned}
&\mathbb{E}\left[\left\|x_i^{k+1}-x_i^{\ast}\right\|^2|\mathcal{F}^k\right]   \leq \left\| x_i^k- x_i^\ast\right\|^2 +2(\lambda^k\mu^k)^2\\
&\qquad + 2\underbrace{(\lambda^k)^2\left\| F_i(x_i^k,v_i^k)-  F_i(x_i^\ast,\bar{x}^{\ast})\right\|^2}_{\rm  Term \:1}\\
&\qquad -\underbrace{2\left\langle x_i^k- x_i^\ast,\, \lambda^k( F_i(x_i^k,v_i^k)-   F_i(x_i^k,\bar{v}^k))\right\rangle}_{\rm Term \: 2}\\
&\qquad -\underbrace{2\left\langle x_i^k- x_i^\ast,\, \lambda^k(  F_i(x_i^k,\bar{v}^k)-  F_i(x_i^\ast,\bar{x}^{\ast}))\right\rangle}_{\rm Term \: 3}.
\end{aligned}
\end{equation}

The three terms on the right hand side of (\ref{eq:x_ast_algo2_2})    can be bounded in a similar way to Theorem \ref{theorem:convergence_algorithm_1}:
\begin{equation}\label{eq:term1_algo2}
\begin{aligned}
{\rm Term \:1}  \leq  12(\lambda^k)^2 C^2 +8(\lambda^k)^2 \tilde{L}^2\|v_i^k-\bar{v}^k\|^2,
\end{aligned}
\end{equation}
\begin{equation}\label{eq:term2_algo2}
\begin{aligned}
&{\rm Term\:2}  \geq -\frac{(\lambda^k)^2\|x_i^k- x_i^\ast\|^2}{\gamma^k},
 -\gamma^k\tilde{L}^2\left\| v_i^k -\bar{v}^k\right\|^2,
\end{aligned}
\end{equation}
\begin{equation}\label{eq:term3_algo2}
\begin{aligned}
{\rm Term \: 3}& =2\lambda^k \left( F_i(x_i^k,\bar{x}^k)-   F_i(x_i^\ast,\bar{x}^{\ast})\right)^T(x_i^k-x_i^{\ast}).
\end{aligned}
\end{equation}

Plugging (\ref{eq:term1_algo2}), (\ref{eq:term2_algo2}), and (\ref{eq:term3_algo2}) into (\ref{eq:x_ast_algo2_2}) yields
\begin{equation}\label{eq:x_ast_2_algo2}
\begin{aligned}
&\mathbb{E}\left[\left\|x_i^{k+1}-x_i^{\ast}\right\|^2|\mathcal{F}^k\right] \\
&\leq \left\| x_i^k- x_i^\ast\right\|^2 +2(\lambda^k\mu^k)^2+ 12(\lambda^k)^2 C^2 \\
&\quad+8(\lambda^k)^2 \tilde{L}^2\|v_i^k-\bar{v}^k\|^2\\
&\quad +\frac{(\lambda^k)^2\|x_i^k- x_i^\ast\|^2}{\gamma^k}
 +\gamma^k\tilde{L}^2\left\| v_i^k -\bar{v}^k\right\|^2\\
&\quad -2 \lambda^k \left( F_i(x_i^k,\bar{x}^k)-  F_i(x_i^\ast,\bar{x}^{\ast})\right)^T(x_i^k-x_i^{\ast}).
\end{aligned}
\end{equation}

Summing (\ref{eq:x_ast_2}) from $i=1$ to $i=m$ yields
\begin{equation}\label{eq:x_ast_2_algo2}
\begin{aligned}
&\mathbb{E}\left[\sum_{i=1}^{m}\left\|x_i^{k+1}-x_i^{\ast}\right\|^2|\mathcal{F}^k\right] \\
&\leq 
\sum_{i=1}^{m}\left\| x_i^k- x_i^\ast\right\|^2 +2m(\lambda^k\mu^k)^2+ 12m(\lambda^k)^2 C^2 \\
&\qquad +8(\lambda^k)^2 \tilde{L}^2\sum_{i=1}^{m}\|v_i^k-\bar{v}^k\|^2\\
&\qquad +\frac{(\lambda^k)^2\sum_{i=1}^{m}\|x_i^k- x_i^\ast\|^2}{\gamma^k}
 +\gamma^k\tilde{L}^2\sum_{i=1}^{m}\left\| v_i^k -\bar{v}^k\right\|^2\\
&\qquad -2 \lambda^k \left(\phi(x^k)- \phi(x^\ast)\right)^T(x^k-x^{\ast}).
\end{aligned}
\end{equation}

Similar to the derivation in Theorem \ref{theorem:convergence_algorithm_1},  we have the following relations from  (\ref{eq:Ev_k_final_algorithm2}) and (\ref{eq:x_ast_2_algo2}) for   $\bv^k=\left[\sum_{i=1}^{m}\|x_i^k-x_i^*\|^2,\ \sum_{i=1}^m \|v_i^k-\bar v^k\|^2\right]^T$:
\begin{equation}\label{eq:stacked_algo2}
\begin{aligned}
\mathbb{E}\left[\bv^{k+1}|\mathcal{F}^k\right]\leq(V^k+A^k)\bv^k-2\lambda^k\Phi^k+B^k,
\end{aligned}
\end{equation}
where

\[
\begin{aligned}
&V^k=\left[\begin{array}{cc}
1 & {\tilde{L}^2}\gamma^k \cr
0& 1-\gamma^k|\rho_2|
\end{array}\right],\\
&A^k=\left[\begin{array}{cc}
\frac{(\lambda^k)^2}{\gamma^k} &
8(\lambda^k)^2\tilde{L}^2\cr
 0& \frac{2(\lambda^k)^2\tilde{L}^2}{\gamma^k|\rho_2|} \end{array}\right],\\
&\Phi^k= \left[\begin{array}{c}
 \left(\phi(x^k)- \phi(x^*)\right)^T(x^k-x^{\ast}) \cr
 0\end{array}\right] ,\\
 & B^k=\left[\begin{array}{c}  2m(\lambda^k\mu^k)^2+12m(\lambda^k)^2 C^2
\cr
\frac{2m(\lambda^k)^2C^2}{\gamma^k|\rho_2|}  +(\gamma^k)^2\|L\|^2 \sum_{i=1}^{m}(\sigma_i^k)^2\end{array}\right].
\end{aligned}
\]

Using Assumption~\ref{assumption:dp-noise} and the conditions of the theorem  $\sum_{k=T}^\infty (\gamma^k)^2<\infty$, $\sum_{k=T}^\infty \frac{(\lambda^k)^2}{\gamma^k}<\infty$, and $\sum_{k=T}^\infty (\lambda^k\mu^k)^2<\infty$, we have that all elements of the matrices of $A^k$ and $B^k$ are summable. Therefore,  we  have $ \sum_{i=1}^{m}\|x_i^k-x_i^*\|^2 $ and $ \sum_{i=1}^m \|v_i^k-\bar v^k\|^2 $ satisfying the conditions of Proposition~\ref{th-main_decreasing} with
$\kappa_1= \tilde{L}^2 $, $\kappa_2=|\rho_2|$, $c^k=2\lambda^k$,
$a^k=\max\{\frac{(\lambda^k)^2}{\gamma^k}, \,
8(\lambda^k)^2\tilde{L}^2,\, \frac{2(\lambda^k)^2\tilde{L}^2}{\gamma^k|\rho_2|} \}$, and $b^k= \max\{ 2m(\lambda^k\mu^k)^2+12m(\lambda^k)^2 C^2,\,
\frac{2m(\lambda^k)^2C^2}{\gamma^k|\rho_2|}  +(\gamma^k)^2\|L\|^2 \sum_{i=1}^{m}(\sigma_i^k)^2\}$.
\end{proof}

\begin{Remark 1}\label{re:variance_gradients}
    Note that different from  \cite{wang2022differentially,lei2022distributed} which deal with  stochastic pseudo-gradients with decreasing variances  (by increasing sample sizes), our Algorithm 2 allows the variance $(\mu^k)^2$ to be constant and even increasing with time. For example, when $\lambda^k$ is set as $\frac{c_1}{1+c_2k}$, the condition   $\sum_{k=T}^\infty (\lambda^k\mu^k)^2<\infty$ in Theorem \ref{theorem:convergence_algorithm_2} can be satisfied for $\mu^k= c_3+c_4k^{\nu}$ with any $0<\nu<0.5$ and positive constants $c_1$, $c_2$, $c_3$, and $c_4$.
\end{Remark 1}

\begin{Remark 1}
Using a reasoning similar to Remark \ref{Re-rate}, we can obtain that the convergence of  $ \sum_{i=1}^{m}\|x_i^{k+1}-x_i^*\|^2$ follows
 \begin{equation}
\begin{aligned}
&\mathbb{E}\left[\sum_{i=1}^{m} \|x_i^{k+1}-x_i^*\|^2|\mathcal{F}^k\right]
\le  (1+a^k)\sum_{i=1}^{m}\|\bar x_i^k -x_i^*\|^2  +  \hat{b}^k,
\end{aligned}
\end{equation}
where all parameters are from Proposition \ref{th-main_decreasing} and $\hat{b}^k=\left(\kappa_1\gamma^k+a^k\right) \sum_{i=1}^m \|v_i^k - \bar v^k\|^2 + b^k$. Given that the stochasticty in $\tilde{F}_i(x_i^k,v_i^k,\xi_i^k)$ only increases the value of  $b^k$ but does not affect its order (still summable), we have that the convergence of all players to the Nash equilibrium is still no slower than the order of $\mathcal{O}(\frac{1}{k})$. Moreover, since the evolution of $v_i^k$ is not affected by the stochasticity in $\tilde{F}_i(x_i^k,v_i^k,\xi_i^k)$ and still follows  (\ref{eq:Ev_k_final}), we have that the decreasing speed of  $\sum_{i=1}^{m}\|v_i^k-  \bar{v}^{k}\|^2 $ still increases with an increase in $|\rho_2|$, which corresponds to the     second largest eigenvalue of $L$. Therefore, the decreasing speed of  $\sum_{i=1}^{m}\|v_i^{k}-  \bar{v}^{k}\|^2 $ to zero increases with an increase in the absolute value of the second largest eigenvalue of $L$ in Assumption \ref{as:L}.
\end{Remark 1}

\subsection{Privacy Analysis for Algorithm 2}

Similar to the privacy analysis in Sec. \ref{se:privacy_Algorithm1}, we can also analyze the strength of differential privacy for Algorithm 2:

\begin{Theorem 1}\label{th:DP_Algorithm2}
Under Assumptions \ref{as:functions}, \ref{as:Lipschitz}, \ref{as:L},  \ref{as:bounded_gradients}, and \ref{As:unbiased}, if  $\{\lambda^k\}$, $\{\gamma^k\}$, and $\{\mu^k\}$ satisfy the conditions in Theorem \ref{theorem:convergence_algorithm_2}, and all elements of $\zeta_i^k$ are drawn independently from  Laplace distribution ${\rm Lap}(\nu^k)$ with $(\sigma_i^k)^2=2(\nu^k)^2$ satisfying Assumption \ref{assumption:dp-noise}, then all players will converge almost surely to the Nash equilibrium. Moreover, 
\begin{enumerate}
\item Algorithm 2 is  $\epsilon$-differentially private with the cumulative privacy budget bounded by $\epsilon\leq \sum_{k=1}^{T_0}\frac{2\bar{C}\lambda^k}{\nu^k}$ for iterations from $k=0$ to $k=T_0$ where $\bar{C}$ is from Assumption \ref{as:bounded_gradients}. And the cumulative privacy budget is always finite for $T_0\rightarrow\infty$  when the sequence  $\{\frac{\lambda^k}{\nu^k}\}$ is summable;
\item Suppose that two non-negative sequences  $\{\nu'^k\}$ and $\{\lambda^k\}$ have a finite sequence-ratio sum $\Phi_{\lambda,\nu'}\triangleq \sum_{k=1}^{\infty}\frac{\lambda^k}{\nu'^k}$. Then setting the Laplace noise parameter $\nu^k$ as $\nu^k=\frac{2 \bar{C}\Phi_{\lambda,\nu'}}{ \epsilon  }\nu'^k$ ensures that Algorithm 2 is $\epsilon$-differentially private with any cumulative privacy budget $\epsilon>0$ even when the number of iterations goes to infinity;
\item In the special case where $\lambda^k=\frac{1}{k}$ and $\gamma^k=\frac{1}{k^{0.9}}$,   setting $\nu^k=\frac{2\bar{C} \Phi}{\epsilon}k^{0.3}$ with  $\Phi\triangleq \sum_{k=1}^{\infty}\frac{1}{k^{1.3}}\approx 3.93$ (which can be verified to satisfy Assumption \ref{assumption:dp-noise}) ensures that  Algorithm 2 is always $\epsilon$-differentially private for any cumulative privacy budget $ \epsilon>0$ even when the number of iterations goes to infinity.
\end{enumerate}
\end{Theorem 1}

\begin{proof}
  The derivation follows the proof of Theorem \ref{th:DP_Algorithm1}, and hence is omitted here.
\end{proof}

\begin{Remark 1}
 Since we use the standard $\epsilon$-differential privacy framework, we characterize the cumulative privacy budget directly.  Under  relaxed (approximate) $\epsilon$-differential privacy frameworks, such as $(\epsilon,\delta)$-differential privacy \cite{kairouz2015composition}, zero-concentrated differential privacy \cite{bun2016concentrated}, or R\'{e}nyi differential privacy \cite{mironov2017renyi},    advanced composition theories in \cite{kairouz2015composition,bun2016concentrated,mironov2017renyi}  can be exploited to characterize the cumulative privacy budget.
\end{Remark 1}

\section{Numerical Simulations}\label{se:simulation}
\begin{figure}
\includegraphics[width=0.45\textwidth]{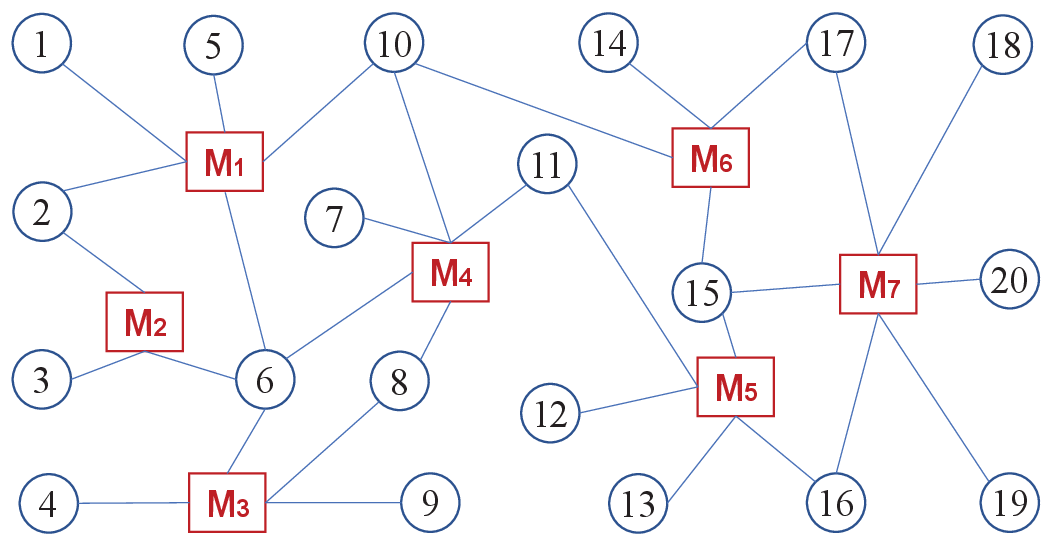}
    \caption{Nash-Cournot game of 20 players (firms) competing over 7 locations (markets). Each firm is represented by a circular and each market is represented by a square. An edge between firm $i$ $(1\leq i\leq 20)$ and market $j$ ($1\leq j\leq 7$) means that firm $i$ participates in market $j$. }
    \label{fig:network}
\end{figure}

 In this section, we evaluate the performance of the  proposed differentially-private distributed Nash-equilibrium computing algorithms   using a networked Nash-Cournot game. More specifically, we consider   $m$ firms  producing a homogeneous commodity competing over $N$ markets, which has been considered recently in    \cite{koshal2016distributed,pavel2019distributed,nguyen2022distributed}. Fig. \ref{fig:network} presents a schematic of the problem involving $N=7$ markets (represented by $M_1,\,\cdots,M_7$) and $m=20$ firms (represented by circles). In the figure, an edge from circle $i$ to $M_j$ means that firm $i$ participates in market $M_j$.

 We consider the  setting where a firm can only see partial decision information of the network. Namely, every firm can only communicate with its immediate neighbors and no central mediator exists which can communicate with all firms. As in \cite{koshal2016distributed,pavel2019distributed}, we allow firms  to communicate with their immediate neighbors   to share their production decisions.  In the considered scenario,  we  use  $x_i\in\mathbb{R}^N$ to represent the amount of firm $i$'  products. Note that a firm $i$  is allowed to participate in $1\leq n_i\leq N$ markets, and if firm $i$ does not participate in market $j$, then the $j$th entry of $x_i$ will be forced to be $0$ all the time. So a firm participating in  $1\leq n_i\leq N$ markets will have $n_i$ non-zero entries in the production vector $x_i$. For the convenience of bookkeeping, we use an adjacency matrix $B_i\in\mathbb{R}^{N\times N}$  to describe the association relationship between firm $i$ and all the markets. More specifically, $B_i$ has zero off-diagonal elements and its  $(j,j)th$ entry is $1$  when firm $i$ participates in market $j$, otherwise, its $(j,j)th$ entry is zero. Every firm $i$ has a maximal capacity for each market $j$ it participates in, which is represented by $C_{ij}$. Denoting $C_i\triangleq [C_{i1},\,\cdots,\,C_{iN}]^T$, we always have $x_i\leq C_i$.
 Represent  $B$ as $B\triangleq [B_1,\,\cdots,\,B_N]$. It can be seen that $Bx\in\mathbb{R}^N=\sum_{i=1}^{N}B_ix_i$ represents the total product supply to all markets, given firm $i$'s production amount $x_i$. As in \cite{pavel2019distributed}, the commodity's price in every  market $M_i$  follows a linear inverse demand function, i.e., it   is a linear function of the total amount of commodity supplied to the market:
 \[
 p_i(x)=\bar{P}_i-\chi_i[Bx]_i,
 \]
 where $\bar{P}_i$ and $\chi_i>0$ are  constants and $[Bx]_i$ denotes the $i$th element of the vector $Bx$. It can be seen that the price decreases with an increase in the amount of supplied commodity.

 We let $p \triangleq [p_1,\,\cdots,\,p_N]^T$ represent the price vector of all markets, which can be verified to satisfy
 \[
 p=\bar{P}-\Xi Bx,
 \]
 where $\bar{P}\triangleq[\bar{P}_1 ,\cdots,\bar{P}_N]^T$ and $\Xi\triangleq{\rm diag}(\chi_1,\,\cdots,\,\chi_N)$. The total payoff of firm $i$ can then be expressed as $p^TB_ix_i$. Firm $i$'s production cost is assumed to be  a strongly convex, quadratic function
 \[
 c_i(x_i)=x_i^TQ_ix_i+q_i^Tx_i,
 \]
 where $Q_i\in\mathbb{R}^{N\times N}$ is a  positive definite matrix and $q_i\in\mathbb{R}^{N}$.

 Therefore, firm $i$'s local objective function, which is determined by its production cost $c_i$ and payoff, is given by
 \[
 f_i(x_i,x)=c_i(x_i)-(\bar{P}-\Xi Bx)^T B_i^T x_i.
 \]
 And it can be verified that the gradient the objective function is
 \[
 F_i(x_i,x)=2Q_ix_i+q_i+B_i^T\Xi B_ix_i-B_i(\bar{P}-\Xi B x).
 \]
 It is clear that both firm $i$'s local objective function and gradient are dependent on other firms' actions.

\begin{figure}
\includegraphics[width=0.45\textwidth]{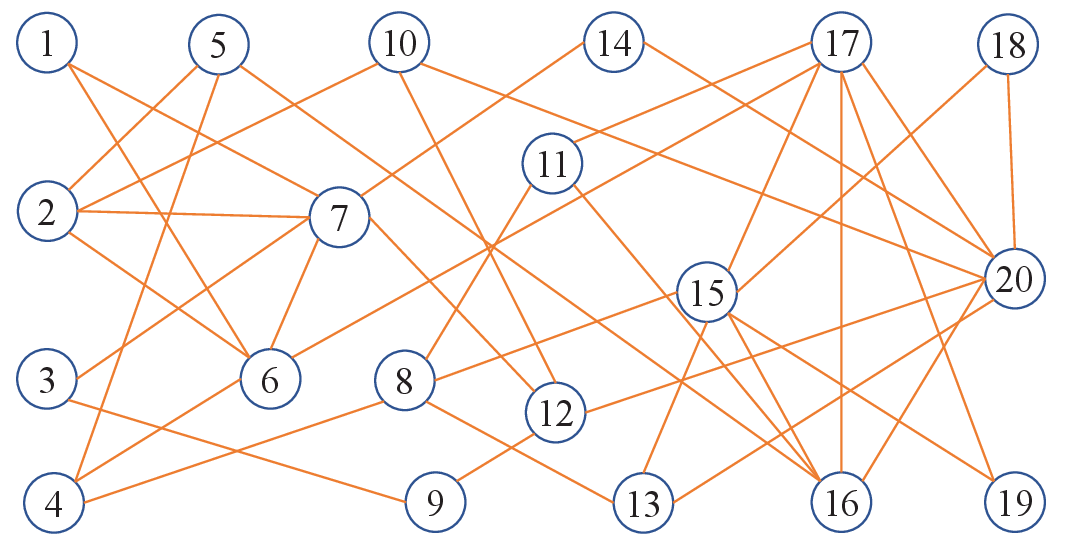}
    \caption{The randomly generated interaction patten of the 20 firms.}
    \label{fig:topology}
\end{figure}

 In the implementation, we consider $N=7$ markets and 20 firms. Since no firm can communicate with all the other firms, we generate  local communication   patterns   randomly, with the interaction graph given in Fig. \ref{fig:topology}. The   maximal capacities for firm $i$ (elements in $C_i$) are randomly selected from the interval $[8,\,10]$. $Q_i$ in the production cost function is set as $\nu I$ with $\nu$ randomly selected from $[1,10]$. $q_i$ in $c_i(x_i)$ is randomly selected from a uniform distribution in $[1,\,2]$. In the price function, $\bar{P}_i$ and $\chi_i$ are randomly chosen from uniform distributions in $[10,\,20]$ and $[1,\,3]$, respectively.

\begin{figure}
\includegraphics[width=0.5\textwidth]{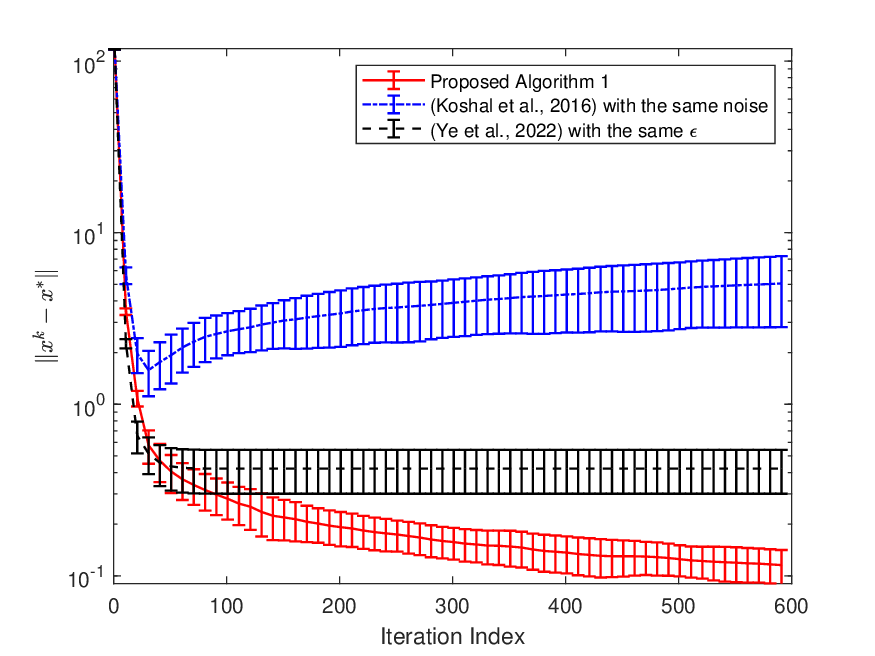}
    \caption{Comparison of Algorithm 1 with the existing distributed Nash-equilibrium computation  algorithm by Koshal et al. in \cite{koshal2016distributed} (under the same noise) and the existing differential-privacy approach for distributed aggregative games by Ye et al. in \cite{ye2021differentially}  (under the same privacy budget $\epsilon$).}
    \label{fig:comparison_algo1}
\end{figure}

  To evaluate the performance of the proposed Algorithm 1, for every firm $i$, we  inject  differential-privacy noise $\zeta_i^k$ in every message it shares in all iterations. Each element of the noise vector follows Laplace  distribution with parameter $\nu^k=1+0.1k^{0.2}$.  We set the stepsize $\lambda^k$ and diminishing sequence $\gamma^k$ as $\lambda^k=\frac{0.1}{1+0.1k}$ and $\gamma^k=\frac{1}{1+0.1k^{0.9}}$, respectively, which satisfy the conditions in Theorem 1 and Theorem 2. In the evaluation, we run our algorithm for 100 times and calculate the average as well as the variance of the gap $\|x^k-x^{\ast}\|$ between generated iterate $x^k$ and the Nash equilibrium $x^{\ast}$  as a function of the iteration index $k$. The result is given by the red curve and error bars in Fig. \ref{fig:comparison_algo1}. For comparison, we also run the existing distributed Nash-equilibrium computation algorithm  proposed by Koshal et al. in \cite{koshal2016distributed} under the same noise, and the existing differential-privacy approach for networked aggregative games proposed by Ye et al. in \cite{ye2021differentially} under the same cumulative privacy budget $\epsilon$. Note that the differential-privacy approach in \cite{ye2021differentially} uses  geometrically decreasing stepsizes (to be summable) to ensure a finite privacy budget, but the fast decreasing stepsize also leads to the loss of guaranteed convergence to the exact Nash equilibrium.   The evolution of the average error/variance of the  approaches in \cite{koshal2016distributed}  and \cite{ye2021differentially} are given by  the blue and black curves/error bars in Fig. \ref{fig:comparison_algo1}, respectively. It is clear that the proposed algorithm has a comparable convergence speed but much better  accuracy.

\begin{figure}
\includegraphics[width=0.5\textwidth]{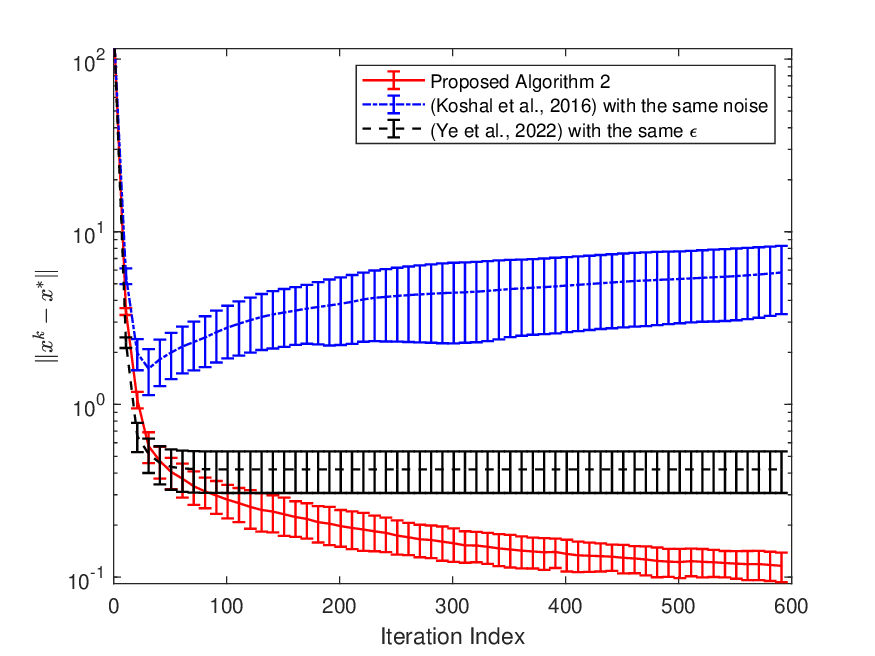}
    \caption{Comparison of Algorithm 2 with the existing stochastic distributed Nash-equilibrium computation algorithm by Koshal et al. in \cite{koshal2016distributed} (under the same noise) and the existing differential-privacy approach for distributed aggregative games by Ye et al. in \cite{ye2021differentially}  (under the same privacy budget $\epsilon$).}
    \label{fig:comparison_algo2}
\end{figure}

Based a similar setup, we also test  the proposed Algorithm 2 when individual players only have access to a stochastic version of the payoff functions and pseudo-gradients. More specifically, we add Gaussian noise of zero mean and unit variance in every dimension of the pseudo-gradient vector $F_i(x_i^k,v_i^k)$. The differential-privacy noise still follows  Laplace  distribution with parameter $\nu^k=1+0.1k^{0.2}$. The stepsize $\lambda^k$ and diminishing sequence $\gamma^k$ are still set as $\lambda^k=\frac{0.1}{1+0.1k}$ and $\gamma^k=\frac{1}{1+0.1k^{0.9}}$, respectively, which satisfy the conditions in Theorem 3 and Theorem 4. The   result is given by the red curve and error bars in Fig. \ref{fig:comparison_algo2}. For comparison, we also run the existing distributed Nash-equilibrium computation algorithm  proposed by Koshal et al. in \cite{koshal2016distributed} under the same noise, and the existing differential-privacy approach for networked aggregative games proposed by Ye et al. in \cite{ye2021differentially} under the same cumulative privacy budget $\epsilon$.    The evolution of the average error/variance of the  approaches in \cite{koshal2016distributed}  and \cite{ye2021differentially} are given by  the blue and black curves/error bars in Fig. \ref{fig:comparison_algo2}, respectively. It is clear that the proposed algorithm has a comparable convergence speed but much better  accuracy.

\section{Conclusions}\label{se:conclusions}
Although differential privacy is becoming the de facto standard for publicly sharing information, its direct incorporation into mediator-free fully distributed aggregative games leads to errors in equilibrium computation due to the need to iteratively and repeatedly inject  independent noises.
This paper proposes a fully distributed Nash-equilibrium computation approach for networked aggregative games  that ensures both accurate convergence to the exact Nash equilibrium and rigorous $\epsilon$-differential privacy with bounded cumulative privacy budget, even when the number of iterations goes to infinity. The simultaneous achievement of both goals is a sharp contrast to existing differential-privacy solutions for  aggregative games that have to trade convergence accuracy for privacy, and to our knowledge, has not been achieved before. The approach can also be extended to stochastic aggregative games and is proven able to ensure both accurate convergence to the Nash equilibrium and rigorous differential privacy, even when every player's  stochastic estimate of the pseudo-gradient is subject to a constant or even  increasing variance.   Numerical simulation   results   confirm that the proposed algorithms have a better accuracy compared with  existing approaches, while maintaining a comparable  convergence speed.


\bibliographystyle{IEEEtran}

\bibliography{reference1}
\vspace{-1.3cm}

\end{document}